\documentclass[%
reprint,
superscriptaddress,
amsmath,amssymb,
aps,
pra,
]{revtex4-1}
\usepackage[utf8]{inputenc}
\usepackage{graphicx}   %include figure files
\usepackage{dcolumn}% Align table columns on decimal point
\usepackage{bm}% bold math
\usepackage[hidelinks]{hyperref}% add hypertext capabilities
\usepackage{braket}
\usepackage{amsmath}
\usepackage{SIunits}
\usepackage{changes}

\usepackage[colorinlistoftodos,prependcaption,textsize=tiny]{todonotes}
\usepackage{multirow}
\begin{document}

\title{An argon ion beam milling process for native $\text{AlO}_\text{x}$ layers enabling coherent superconducting contacts}

\author{Lukas Gr\"unhaupt}
\affiliation{Physikalisches Institut, Karlsruhe Institute of Technology, 76131 Karlsruhe, Germany}

\author{Uwe von L\"upke}
\affiliation{Physikalisches Institut, Karlsruhe Institute of Technology, 76131 Karlsruhe, Germany}

\author{Daria Gusenkova}
\affiliation{Physikalisches Institut, Karlsruhe Institute of Technology, 76131 Karlsruhe, Germany}
\affiliation{Russian Quantum Center, National University of Science and Technology MISIS, 119049 Moscow, Russia}

\author{Sebastian T. Skacel}
\affiliation{Physikalisches Institut, Karlsruhe Institute of Technology, 76131 Karlsruhe, Germany}

\author{Nataliya Maleeva}
\affiliation{Physikalisches Institut, Karlsruhe Institute of Technology, 76131 Karlsruhe, Germany}

\author{Steffen Schl\"or}
\affiliation{Physikalisches Institut, Karlsruhe Institute of Technology, 76131 Karlsruhe, Germany}

\author{Alexander Bilmes}
\affiliation{Physikalisches Institut, Karlsruhe Institute of Technology, 76131 Karlsruhe, Germany}

\author{Hannes Rotzinger}
\affiliation{Physikalisches Institut, Karlsruhe Institute of Technology, 76131 Karlsruhe, Germany}

\author{Alexey V. Ustinov}
\affiliation{Physikalisches Institut, Karlsruhe Institute of Technology, 76131 Karlsruhe, Germany}
\affiliation{Russian Quantum Center, National University of Science and Technology MISIS, 119049 Moscow, Russia}

\author{Martin Weides}
\affiliation{Physikalisches Institut, Karlsruhe Institute of Technology, 76131 Karlsruhe, Germany}
\affiliation{Physikalisches Institut, Johannes Gutenberg University Mainz, 55128 Mainz, Germany}

\author{Ioan M. Pop}
\email{ioan.pop@kit.edu}
\affiliation{Physikalisches Institut, Karlsruhe Institute of Technology, 76131 Karlsruhe, Germany}

\date{\today}% It is always \today, today,
%-----------------------------------------------------------------------------

\begin{abstract}
We present an argon ion beam milling process to remove the native oxide layer forming on aluminum thin films due to their exposure to atmosphere in between lithographic steps. Our cleaning process is readily integrable with conventional fabrication of Josephson junction quantum circuits. From measurements of the internal quality factors of superconducting microwave resonators with and without contacts, we place an upper bound on the residual resistance of an ion beam milled contact of \unit{50}{\milli\ohm\cdot\micro\meter\squared} at a frequency of \unit{4.5}{\giga\hertz}. Resonators for which only 6\% of the total foot-print was exposed to the ion beam milling, in areas of low electric and high magnetic field, showed quality factors above $10^6$ in the single photon regime, and no degradation compared to single layer samples. We believe these results will enable the development of increasingly complex superconducting circuits for quantum information processing.   
\end{abstract}

%-----------------------------------------------------------------------------

\maketitle

%-----------------------------------------------------------------------------
% Introduction
The research field of superconducting quantum electronics has been developing at an accelerated pace for the last two decades, and it is now one of the leading candidates for the implementation of quantum mechanical computational machines which could eventually outperform classical computers \cite{devoret_superconducting_2013}. On the path to achieving this scientific landmark, microelectronic quantum circuits are required to become increasingly complex, to implement an ever growing set of functionalities, such as: fast single and multiple qubit operations \cite{plantenberg_demonstration_2007, barends_superconducting_2014}, quantum non-demolition readout \cite{wallraff_strong_2004, castellanos-beltran_widely_2007, vijay_observation_2011, abdo_josephson_2011}, remote qubit entanglement \cite{riste_deterministic_2013, steffen_deterministic_2013, chow_implementing_2014, roch_observation_2014} and qubit-qubit interactions \cite{berkley_entangled_2003, steffen_measurement_2006, majer_coupling_2007}, or autonomous feedback \cite{shankar_2013_autonomously, geerlings_demonstrating_2013}.  Not only are these goals challenging by themselves, but it is paramount that they are achieved without compromising on the quantum coherence of the device.

Several quantum circuit integration approaches are currently pursued with promising results. Flip-chip strategies \cite{minev_planar_2016, brecht_demonstration_2015} or complex 2.5D circuit designs \cite{kelly_state_2015, riste_detecting_2015} have recently shown coherence comparable with state of the art single devices \cite{paik_observation_2011,rigetti_superconducting_2012,barends_coherent_2013}. Their fabrication often requires several lithography steps, involving different clean-room technologies. One of the challenges of integrating different microelectronic fabrication layers \cite{sandberg_etch_2012,vissers_identifying_2012,braumuller_multiphoton_2015}, is to obtain not only a very good galvanic contact, but also a very high quality factor at microwave frequencies.

Aluminum is one of the most widely used materials for superconducting quantum electronics, thanks to the controllable and convenient growth of the oxide barrier between the electrodes of Josephson junctions, and its relatively low surface dielectric loss tangent \cite{oconnell_microwave_2008,wang_surface_2015}. However, aluminum also forms an insulating oxide when exposed to atmosphere, which has to be removed prior to contacting different lithographic layers.

In this letter we present an argon ion beam milling process to remove the native aluminum oxide, which enables the fabrication of state of the art coherent devices. We show that overlap contacts obtained using ion beam milling did not cause any measurable degradation in quality factor compared to a continuous metallic film, when embedded into microwave resonators with internal quality factors on the order of $10^6$ in the single photon regime.

\begin{figure}[!t]
\begin{center}
\includegraphics[width=.9\columnwidth]{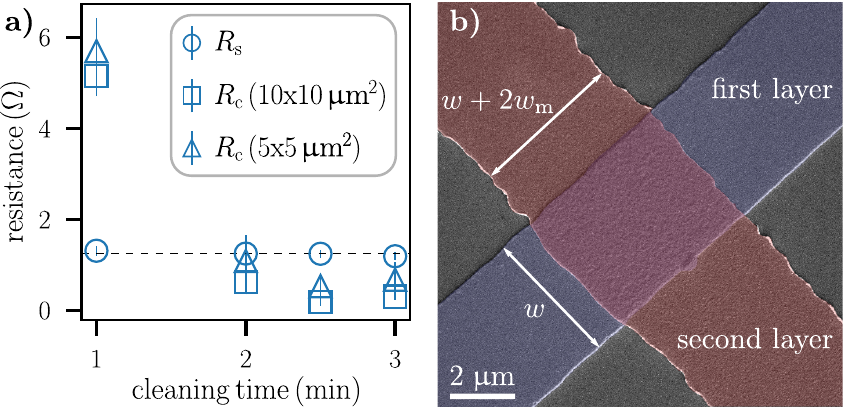}
\caption{\textbf{a)} Measured DC contact resistance $R_{\mathrm{c}}$ at room temperature as a function of cleaning time for $5\times5$ and \unit{10\times10}{\micro\meter\squared} overlaps. For times longer than \unit{2}{\minute}, $R_{\mathrm{c}}$ is below the sheet resistance $R_{\mathrm{s}}$ of a single layer aluminum film. Details on the contact resistance measurement are provided in the supplementary material.  \textbf{b)} False colored SEM image of a contact area after \unit{3}{\minute} of cleaning and deposition of the second aluminum layer. The second (red) layer shows rough edges due to the aggressive cleaning step performed prior to metal deposition, and a widening of the strip by $2w_{\mathrm{m}}$. From multiple measurements on SEM images we observe that the widening of the strip does not change for milling times between 1 and \unit{3}{\minute}, and $2w_{\mathrm{m}}$ can be as large as \unit{1}{\micro\meter}. The difference in resist height before and after milling was measured to be less than \unit{20}{\nano\meter}.}
\label{fig:1}
\end{center}
\end{figure}

\begin{figure*}[!t]
\begin{center}
\includegraphics[scale=.9]{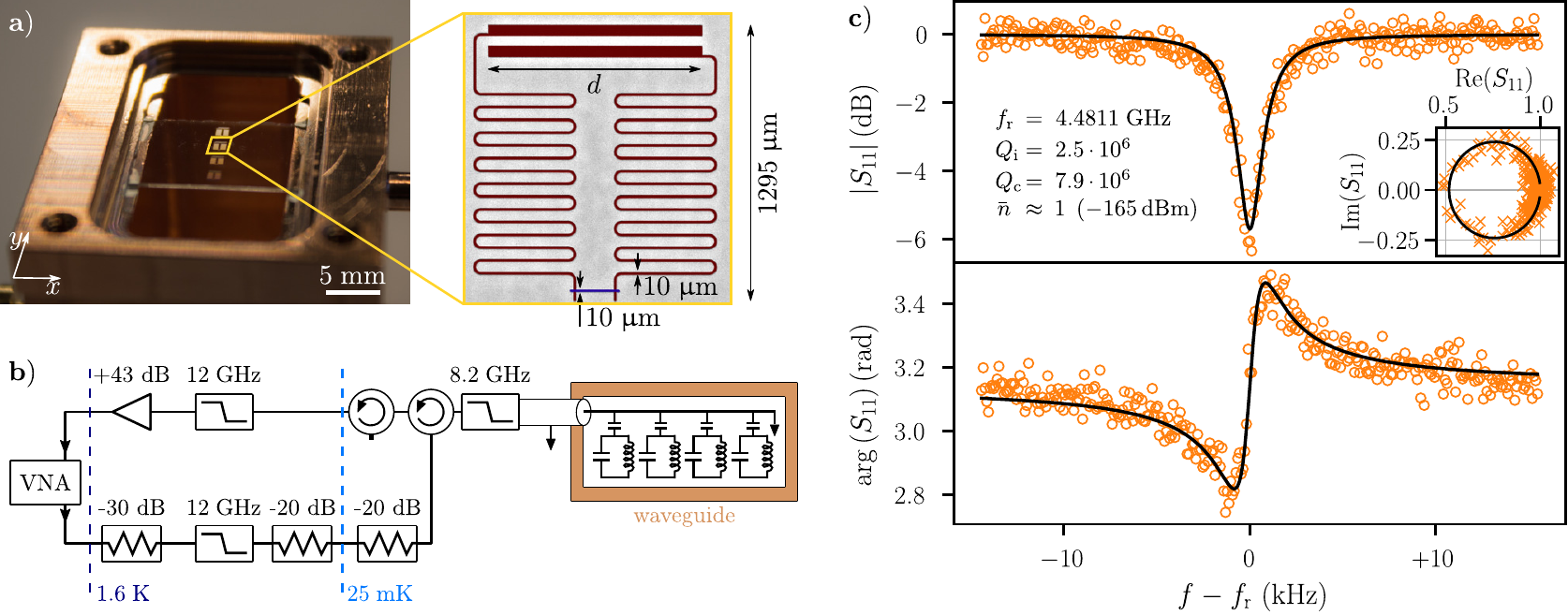}
\caption{\textbf{a)} Photograph of a sample mounted in the copper waveguide sample holder. Each chip holds four resonators. Their dipole moment couples to the $\mathrm{TE}_{\mathrm{10}}$ mode of the waveguide (E-field parallel to the x direction of the indicated coordinate system). The inset shows an optical microscope image of a resonator in false color. To test the coherent properties of the cleaned contacts, the meander (red) is interrupted in the middle and closed in a second lithographic step by an aluminum thin film of the same width, that we call \textit{bridge} (blue). Varying the length $d$ of the capacitor sweeps the resonant frequencies. All remaining parameters of the design are nominally identical between resonators on one chip.  \textbf{b)} Schematic of the cryogenic measurement setup. A reflection measurement with a vector network analyzer (VNA) characterizes the resonator response. The waveguide is thermally anchored to the mixing chamber plate of a commercial dilution refrigerator. All input and output lines are interrupted by commercial and custom made low pass filters providing at least -30\,dB of filtering above \unit{9}{\giga\hertz}. Including cables, the total attenuation on the input lines is -70\,dB. \textbf{c)} Typical measured and fitted (black lines) reflection data of a resonator at an estimated drive power of -165\,dBm at the waveguide input, corresponding to an average number of photons  $\bar{n} \approx 1$  circulating in the resonator. The fitted values for the resonant frequency $f_{\mathrm{r}}$, the internal quality factor $Q_{\mathrm{i}}$, and the coupling quality factor $Q_{\mathrm{c}}$ are indicated in the plot.}
\label{fig:2}
\end{center}
\end{figure*}

We perform the argon ion beam milling using the Kaufman ion source connected to the load lock of a $\mathrm{Plassys}^{\mathrm{TM}}$ MEB 550 S shadow evaporation machine at a base pressure in the range of $10^{-7}$\,mbar immediately before the deposition of aluminum thin films. The parameters of the ion source during cleaning are set as follows: 4 sccm argon-gas at a beam voltage of \unit{400}{\volt}, an accelerating voltage of \unit{90}{V} and an ion current of \unit{15}{\milli\ampere}. 
Between the end of the milling process and the opening of the shutter for the aluminum deposition the time interval is approximately \unit{300}{\second}. 
The rate of the aluminum deposition is \unit{0.2}{\nano\meter\per\second}.
All samples are fabricated using an optical lithography lift-off technique on double-side polished \unit{330}{\micro\meter} thick c-plane sapphire wafers.   

We calibrate the duration of the milling process by measuring the decrease in DC resistance of overlap contacts, as shown in Fig.~\ref{fig:1}a. After \unit{2}{\minute} of milling the DC contact resistance $R_{\mathrm{c}}$ is smaller than the sheet resistance $R_{\mathrm{s}}$ of a single-layer aluminum film, which defines the measurable upper bound for $R_{\mathrm{c}}$ in our setup. Figure~\ref{fig:1}b shows a SEM image where we can observe the effect of the aggressive cleaning step on the patterned resist: a widening of the strips by $2w_{\mathrm{m}} \lesssim \unit{1}{\micro\meter}$ together with a roughening of the edges. By completely milling aluminum films with and without exposure to atmosphere, we estimate a ratio of 1:10 for the milling rates of native aluminum oxide and pure aluminum. Given the thicknesses of the native oxide layer (\unit{1-2}{\nano\meter}) and the underlying pure aluminum thin film (\unit{35}{\nano\meter}), we expect comparable milling times for their removal. Therefore, it is crucial not to overetch the overlapping area of the contacts. Figure~\ref{fig:1}b shows that after \unit{3}{\minute} of cleaning, one minute longer than what is needed for a negligible DC resistance, the first aluminum electrode is still continuous, illustrating the robustness of the process with respect to possible ion beam milling inhomogeneities.

%-----------------------------------------------------------------------------
% Resonator measurements

To measure the coherence of the contacts between lithographic layers we fabricate superconducting resonators with and without overlap contacts and compare their quality factors. All lithographic layers are done in conventional lift-off technique.  Figure~\ref{fig:2}a shows a picture of a \unit{10 \times 15}{\milli\meter\squared} sapphire chip with four resonators mounted in a 3D copper waveguide sample holder, following a design that was recently used to perform simultaneous readout of fluxonium qubits \cite{kou_simultaneous_2017}. The sample holder has a pass band of approximately \unit{1.5}{\giga\hertz} starting from the cutoff frequency of the waveguide at \unit{5.8}{\giga\hertz}. Inside the band, the reflection from the waveguide to the \unit{50}{\ohm} coaxial cables of our measurement setup is below $-12\,\mathrm{dB}$. A copper cap closes and shorts the waveguide at a distance of \unit{8}{\milli\meter} (approximately $\lambda/4$ for frequencies in the bandwidth) from the sapphire chip. Silver paste fixes the sapphire chip to the waveguide body and an indium wire seal ensures good electrical contact as well as tight sealing between the waveguide body and the cap. Two shields machined from a copper/aluminum sandwich and $\mu$-metal around the closed waveguide sample holder provide IR radiation \cite{barends_minimizing_2011} and magnetic shielding (see supplementary material). The entire assembly is thermally anchored to the base plate of a commercial dilution refrigerator at \unit{20}{\milli\kelvin}.  

\begin{figure}[!th]
\begin{center}
\includegraphics[width=0.9\columnwidth]{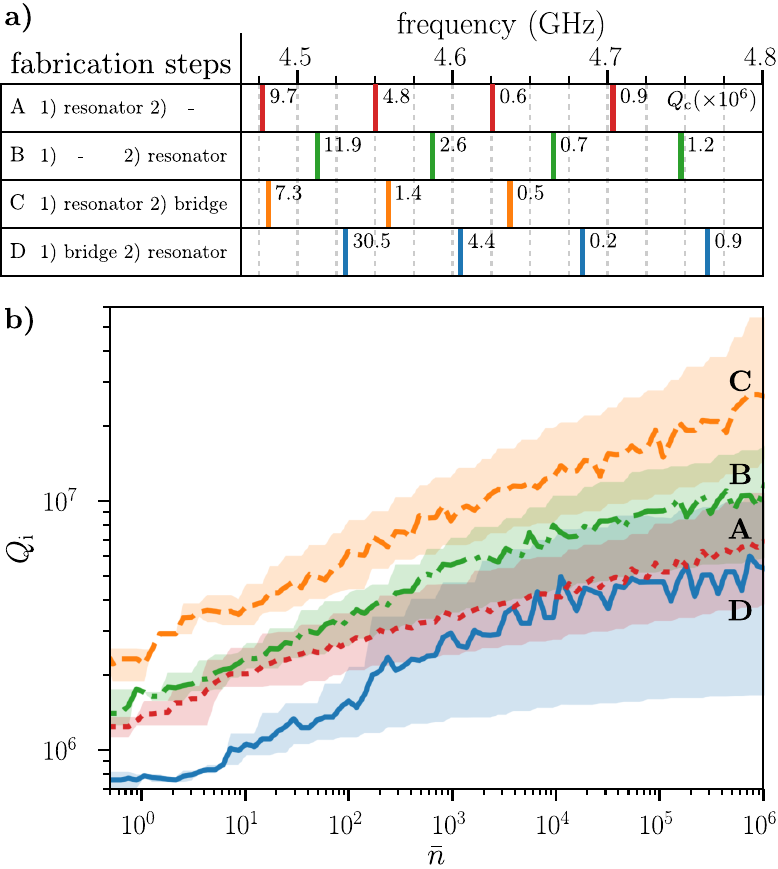}
\caption{\textbf{a)} Overview of all measured samples. The left column shows the order of fabrication, the right column indicates the resonant frequencies and the corresponding coupling quality factors $Q_{\mathrm{c}}$ in millions. We achieve these weak coupling values by designing resonant frequencies below the cutoff frequency of the waveguide (\unit{5.8}{\giga\hertz}). On each sample the resonators are equally spaced in frequency, approximately \unit{75}{\mega\hertz} apart. For an extensive parameter list see the supplementary material. The first aluminum layer is deposited with a thickness of \unit{35}{\nano\meter} and the second layer with a thickness of \unit{50}{\nano\meter}. One of the meanders on sample C is interrupted, leaving only three functional resonators. \textbf{b)} Quality factors of the four investigated samples as a function of the average number of circulating photons $\bar{n}$. Solid lines indicate the mean internal quality factor $Q_{\mathrm{i}}$ of all fitted resonators of each sample, while the shaded areas show the spread between the highest and the lowest measured $Q_{\mathrm{i}}$.}
\label{fig:3}
\end{center}
\end{figure}

In the inset of Fig.~\ref{fig:2}a we show an optical microscope image of one of the measured resonators. For all resonators, the length of the meandering inductor $l$ is \unit{16}{\milli\meter} and its width $w$ is \unit{10}{\micro\meter}, while the capacitor length $d$ takes the values 1000, 950, 900, and \unit{850}{\micro\meter}, which distribute the resonant frequencies in a range of \unit{300}{\mega\hertz} around \unit{4.6}{\giga\hertz}. We deliberately design these frequencies below the cutoff frequency of the waveguide to decouple the resonators from the microwave environment and achieve coupling quality factors $Q_{\mathrm{c}}$ in the range of $10^6$ (see Fig.~\ref{fig:3}a). To test the quality factor of the argon ion milled overlap contacts, for half of the measured resonators the aluminum film of the meander is interrupted in the middle, and reconnected in a second lithographic step using a strip of the same width that we call \textit{bridge}. 

Figure~\ref{fig:2}b shows a schematic of our cryogenic measurement setup. A vector network analyzer (VNA) measures the complex response of the resonators. The input signal is in total attenuated by -100 dB, -30 dB at room temperature and -70 dB distributed at different temperature stages of the cryostat, including the attenuation of the resistive coaxial microwave cables. Two cryogenic circulators provide signal routing and isolation on the output line, respectively. A commercial high electron mobility transistor amplifier on the \unit{1.6}{\kelvin} stage of the cryostat amplifies the outgoing signal by +43 dB. At room temperature a second commercial amplifier adds +60 dB to the signal. 

Figure~\ref{fig:2}c shows a typical measured and fitted $S_{11}$ resonator response at an input power of -165 dBm corresponding to an average number of photons circulating in the resonator of $\bar{n} = 4 P_{\mathrm{in}} Q_{\mathrm{tot}}^2 / \hbar \omega_{\mathrm{r}}^2 Q_{\mathrm{c}} \approx 1$. We use a circle fit routine in the complex plane to extract the quality factor and the resonant frequency \cite{probst_efficient_2015}.

\begin{figure}[t!]
\begin{center}
\includegraphics[width=0.9\columnwidth]{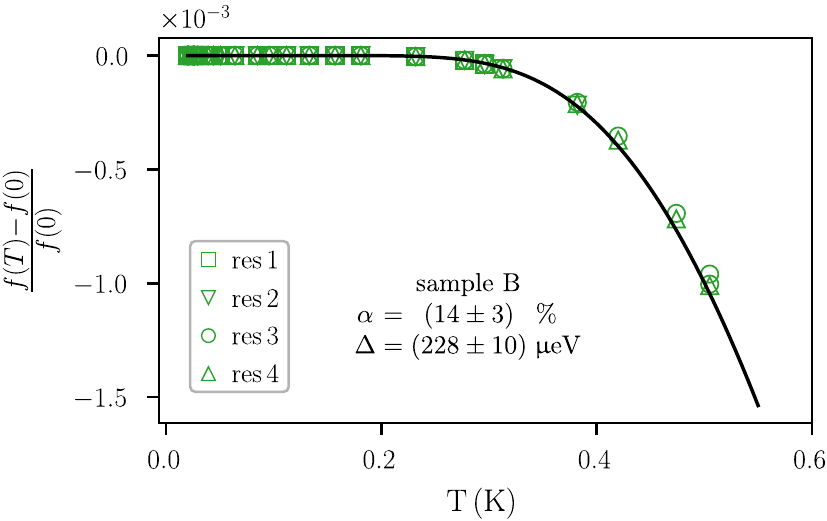}
\caption{Measurement of the kinetic inductance fraction $\alpha$. The symbols show the relative shift of the resonant frequency of the four resonators on sample B as a function of temperature. The data of each resonator is fitted using Eq. \eqref{eq:f_of_T}. The black line shows the model using the average values of the four individual fits: $\alpha = (14\pm 3)$\% and $\Delta = \unit{228\pm 10}{\micro\electronvolt}$.}
\label{fig:4}
\end{center}
\end{figure}

For each sample, Fig.~\ref{fig:3} gives an overview of the fabrication sequence, the measured internal quality factors $Q_{\mathrm{i}}$, the resonant frequencies, and corresponding coupling quality factors $Q_{\mathrm{c}}$. Notice that the frequencies of the resonators on samples A and C are significantly lower than those of samples B and D. This can be explained by the fact that the entire resonator, except the bridge region, is deposited in lithography step 1 for samples A and C, and in step 2 for B and D. The frequency shift between the two groups of samples is caused by the ion milling step which increases the meander width (see Fig.~\ref{fig:1}) by $2w_{\mathrm{m}}$, effectively reducing the number of squares, $l/(w+2w_{\mathrm{m}})$, for samples B and D, and thereby decreasing the kinetic inductance. The observed shift of approximately \unit{40}{\mega\hertz} could be explained by a widening of the strip $2w_{\mathrm{m}}$ on the order of \unit{1}{\micro\meter}, which is consistent with the values indicated in Fig.~\ref{fig:1}b. Additionally, the milling lowers the geometric inductance of the meander, and also increases the capacitance by reducing the distance between the capacitor pads. However, these two modifications to the resonator geometry will only be on the order of 1\%, the resulting changes in resonant frequency will have opposite signs, and they can therefore be neglected. We estimate that the difference between the film thickness of samples A, C (\unit{35}{\nano\meter}) and B, D (\unit{50}{\nano\meter}) will result in a $\sim$10 \% change of the kinetic inductance fraction $\alpha$~\cite{gao_physics_2008}. For sample B we measured $\alpha=0.14$ (see Fig.~\ref{fig:4}), which implies that the frequency shift between samples A, C and B, D, due to the change in kinetic inductance fraction should be less than \unit{5}{\mega\hertz}.

The measured frequency difference between resonators fabricated in the same lithographic step without, A and B, and with a contact bridge, C and D, respectively, is only on the order of a few \unit{}{\mega\hertz}. Surprisingly, the resonant frequencies of samples with a contact bridge are all higher than the frequencies of the corresponding single layer resonators, indicating a negligible contribution from the kinetic inductance of the overlap contacts. The shift to higher frequencies for resonators on sample C compared to sample A could again be explained by a widening of the bridge during the argon ion milling, consistent with the arguments presented in the previous paragraph. Finally, for sample D, we expect smaller frequencies compared to sample B, however, they are measured to be significantly higher. This shift, observed for samples where the entire area of the resonator was subjected to the ion milling, could also arise from random fluctuations of the width $w$ of structures fabricated in different positions on the wafer. Possible causes for this variations include a non-uniform ion beam profile or inhomogeneities in the UV-beam exposure over the two inch diameter of the wafer. 

The solid lines in Fig.~\ref{fig:3}b show the mean $Q_{\mathrm{i}}$ of all resonators for each sample as a function of the average number of circulating photons $\bar{n}$. The spread between the highest and lowest $Q_{\mathrm{i}}$ of each sample is indicated by the shaded area. We would like to emphasize that the single-layer samples A and B, and sample C, where the cleaning process is only applied to the connecting bridge, show internal quality factors larger than $10^6$ in the single photon regime. Remarkably, we measure the highest average $Q_{\mathrm{i}}$ on the resonators of sample C which include an ion beam milled contact bridge. From the mean internal quality factors of samples with a bridge (C, D) we extract an upper bound on the residual resistance of the contacts of $\unit{50}{\milli\ohm\cdot\micro\meter\squared}$ (see supplementary material). From 3D finite element simulations we extract a participation ratio of the metal-substrate and metal-air interfaces of the resonator of $p = 1.8 \cdot 10^{-4}$, which is smaller than in coplanar geometries, due to the larger mode volume of the waveguide sample holder. This allows us to extract a surface dielectric loss tangent $\tan \delta = (p\, \bar{Q_{\mathrm{i}}})^{-1} = 4 \cdot 10^{-3}$, which is in the range of commonly reported values~\cite{wang_surface_2015}.

A potential inhomogeneity of the argon ion beam could have caused stronger milling of sample D and a degradation of the substrate \cite{dunsworth_characterization_2017}, thereby explaining the lower quality factors of all resonators of sample D, compared to those of sample B.

To measure the kinetic inductance fraction $\alpha$ of sample B, for which the entire surface of the resonators was subjected to the milling, we measure the temperature dependence of the resonant frequencies. The measured change of the resonant frequency as a function of temperature (see Fig.~\ref{fig:4}) is modeled using the following equation, 
\begin{equation}
df(T)/f_{\mathrm{r}} = -\alpha / 2 \sqrt{\pi \Delta / 2 k_{\mathrm{B}} T} \exp(-\Delta / k_{\mathrm{B}} T),
\label{eq:f_of_T}
\end{equation} 
where the kinetic inductance fraction $\alpha = L_{\mathrm{kin}} / L_{\mathrm{tot}}$ and the superconducting gap $\Delta$ are used as fit parameters.

The fit is performed for all four resonators of sample B individually. Taking the average of all fitted parameters yields mean values of $\alpha = (14 \pm 3)$\%, and $\Delta = \unit{(228 \pm 10)}{\micro\electronvolt}$ which corresponds to a BCS critical temperature of $(1.5 \pm \unit{0.1)}{\kelvin}$. Therefore, we do not observe any change in the intrinsic properties of the aluminum thin film deposited after the ion beam milling process compared to standard aluminum thin films \cite{gao_physics_2008}. 

We have demonstrated an argon ion beam milling process for the removal of the native oxide layer forming on aluminum thin films. Measurements of superconducting microwave resonators in the single photon regime show no degradation of $Q_{\mathrm{i}}$ at a level of $10^6$ when the milling process is used on a small area of the superconducting circuit. Very recently, similar bounds on coherence were reported for overlap Josephson junctions \cite{wu_overlap_2017} and contacts \cite{dunsworth_characterization_2017}. If the milling is performed on the entire area of the resonator it induces at most a factor of two degradation in $Q_{\mathrm{i}}$.
These results enable the development of increasingly complex superconducting circuit designs with several interconnected lithographic layers, without compromising their coherence properties, thereby opening the way to the integration of very different and often complementary quantum systems such as Josephson junctions and superconducting high kinetic inductance nanowires \cite{mooij_superconducting_2006, astafiev_coherent_2012}, or mesoscopic semiconductor structures \cite{cleuziou_carbon_2006,larsen_semiconductor-nanowire-based_2015, de_lange_realization_2015}. 

We are grateful to L. Radtke, A. Lukashenko, and P. Winkel for technical support. Facilities use was supported by the KIT Nanostructure Service Laboratory (NSL). Funding was provided by the Alexander von Humboldt foundation in the framework of a Sofja Kovalevskaja award endowed by the German Federal Ministry of Education and Research. IMP and MW acknowledge partial financial support from the KIT Young Investigator Network (YIN). MW acknowledges funding by the European Research Council (CoG 648011). DG acknowledges support from the Karlsruhe House of Young Scientists (KHYS). AU acknowledges partial support from the Russian Federation Ministry of Education and Science (NUST MISIS contract no. K2-2016-063).

\bibliography{argon_ion_milling_references}

\begin{thebibliography}{39}%
\makeatletter
\providecommand \@ifxundefined [1]{%
 \@ifx{#1\undefined}
}%
\providecommand \@ifnum [1]{%
 \ifnum #1\expandafter \@firstoftwo
 \else \expandafter \@secondoftwo
 \fi
}%
\providecommand \@ifx [1]{%
 \ifx #1\expandafter \@firstoftwo
 \else \expandafter \@secondoftwo
 \fi
}%
\providecommand \natexlab [1]{#1}%
\providecommand \enquote  [1]{``#1''}%
\providecommand \bibnamefont  [1]{#1}%
\providecommand \bibfnamefont [1]{#1}%
\providecommand \citenamefont [1]{#1}%
\providecommand \href@noop [0]{\@secondoftwo}%
\providecommand \href [0]{\begingroup \@sanitize@url \@href}%
\providecommand \@href[1]{\@@startlink{#1}\@@href}%
\providecommand \@@href[1]{\endgroup#1\@@endlink}%
\providecommand \@sanitize@url [0]{\catcode `\\12\catcode `\$12\catcode
  `\&12\catcode `\#12\catcode `\^12\catcode `\_12\catcode `\%12\relax}%
\providecommand \@@startlink[1]{}%
\providecommand \@@endlink[0]{}%
\providecommand \url  [0]{\begingroup\@sanitize@url \@url }%
\providecommand \@url [1]{\endgroup\@href {#1}{\urlprefix }}%
\providecommand \urlprefix  [0]{URL }%
\providecommand \Eprint [0]{\href }%
\providecommand \doibase [0]{http://dx.doi.org/}%
\providecommand \selectlanguage [0]{\@gobble}%
\providecommand \bibinfo  [0]{\@secondoftwo}%
\providecommand \bibfield  [0]{\@secondoftwo}%
\providecommand \translation [1]{[#1]}%
\providecommand \BibitemOpen [0]{}%
\providecommand \bibitemStop [0]{}%
\providecommand \bibitemNoStop [0]{.\EOS\space}%
\providecommand \EOS [0]{\spacefactor3000\relax}%
\providecommand \BibitemShut  [1]{\csname bibitem#1\endcsname}%
\let\auto@bib@innerbib\@empty
%</preamble>
\bibitem [{\citenamefont {Devoret}\ and\ \citenamefont
  {Schoelkopf}(2013)}]{devoret_superconducting_2013}%
  \BibitemOpen
  \bibfield  {author} {\bibinfo {author} {\bibfnamefont {M.~H.}\ \bibnamefont
  {Devoret}}\ and\ \bibinfo {author} {\bibfnamefont {R.~J.}\ \bibnamefont
  {Schoelkopf}},\ }\href {\doibase 10.1126/science.1231930} {\bibfield
  {journal} {\bibinfo  {journal} {Science}\ }\textbf {\bibinfo {volume}
  {339}},\ \bibinfo {pages} {1169} (\bibinfo {year} {2013})}\BibitemShut
  {NoStop}%
\bibitem [{\citenamefont {Plantenberg}\ \emph {et~al.}(2007)\citenamefont
  {Plantenberg}, \citenamefont {{de Groot}}, \citenamefont {Harmans},\ and\
  \citenamefont {Mooij}}]{plantenberg_demonstration_2007}%
  \BibitemOpen
  \bibfield  {author} {\bibinfo {author} {\bibfnamefont {J.~H.}\ \bibnamefont
  {Plantenberg}}, \bibinfo {author} {\bibfnamefont {P.~C.}\ \bibnamefont {{de
  Groot}}}, \bibinfo {author} {\bibfnamefont {C.~J. P.~M.}\ \bibnamefont
  {Harmans}}, \ and\ \bibinfo {author} {\bibfnamefont {J.~E.}\ \bibnamefont
  {Mooij}},\ }\href {\doibase 10.1038/nature05896} {\bibfield  {journal}
  {\bibinfo  {journal} {Nature}\ }\textbf {\bibinfo {volume} {447}},\ \bibinfo
  {pages} {836} (\bibinfo {year} {2007})}\BibitemShut {NoStop}%
\bibitem [{\citenamefont {Barends}\ \emph {et~al.}(2014)\citenamefont
  {Barends}, \citenamefont {Kelly}, \citenamefont {Megrant}, \citenamefont
  {Veitia}, \citenamefont {Sank}, \citenamefont {Jeffrey}, \citenamefont
  {White}, \citenamefont {Mutus}, \citenamefont {Fowler}, \citenamefont
  {Campbell}, \citenamefont {Chen}, \citenamefont {Chen}, \citenamefont
  {Chiaro}, \citenamefont {Dunsworth}, \citenamefont {Neill}, \citenamefont
  {O'Malley}, \citenamefont {Roushan}, \citenamefont {Vainsencher},
  \citenamefont {Wenner}, \citenamefont {Korotkov}, \citenamefont {Cleland},\
  and\ \citenamefont {Martinis}}]{barends_superconducting_2014}%
  \BibitemOpen
  \bibfield  {author} {\bibinfo {author} {\bibfnamefont {R.}~\bibnamefont
  {Barends}}, \bibinfo {author} {\bibfnamefont {J.}~\bibnamefont {Kelly}},
  \bibinfo {author} {\bibfnamefont {A.}~\bibnamefont {Megrant}}, \bibinfo
  {author} {\bibfnamefont {A.}~\bibnamefont {Veitia}}, \bibinfo {author}
  {\bibfnamefont {D.}~\bibnamefont {Sank}}, \bibinfo {author} {\bibfnamefont
  {E.}~\bibnamefont {Jeffrey}}, \bibinfo {author} {\bibfnamefont {T.~C.}\
  \bibnamefont {White}}, \bibinfo {author} {\bibfnamefont {J.}~\bibnamefont
  {Mutus}}, \bibinfo {author} {\bibfnamefont {A.~G.}\ \bibnamefont {Fowler}},
  \bibinfo {author} {\bibfnamefont {B.}~\bibnamefont {Campbell}}, \bibinfo
  {author} {\bibfnamefont {Y.}~\bibnamefont {Chen}}, \bibinfo {author}
  {\bibfnamefont {Z.}~\bibnamefont {Chen}}, \bibinfo {author} {\bibfnamefont
  {B.}~\bibnamefont {Chiaro}}, \bibinfo {author} {\bibfnamefont
  {A.}~\bibnamefont {Dunsworth}}, \bibinfo {author} {\bibfnamefont
  {C.}~\bibnamefont {Neill}}, \bibinfo {author} {\bibfnamefont
  {P.}~\bibnamefont {O'Malley}}, \bibinfo {author} {\bibfnamefont
  {P.}~\bibnamefont {Roushan}}, \bibinfo {author} {\bibfnamefont
  {A.}~\bibnamefont {Vainsencher}}, \bibinfo {author} {\bibfnamefont
  {J.}~\bibnamefont {Wenner}}, \bibinfo {author} {\bibfnamefont {A.~N.}\
  \bibnamefont {Korotkov}}, \bibinfo {author} {\bibfnamefont {A.~N.}\
  \bibnamefont {Cleland}}, \ and\ \bibinfo {author} {\bibfnamefont {J.~M.}\
  \bibnamefont {Martinis}},\ }\href {\doibase 10.1038/nature13171} {\bibfield
  {journal} {\bibinfo  {journal} {Nature}\ }\textbf {\bibinfo {volume} {508}},\
  \bibinfo {pages} {500} (\bibinfo {year} {2014})}\BibitemShut {NoStop}%
\bibitem [{\citenamefont {Wallraff}\ \emph {et~al.}(2004)\citenamefont
  {Wallraff}, \citenamefont {Schuster}, \citenamefont {Blais}, \citenamefont
  {Frunzio}, \citenamefont {Huang}, \citenamefont {Majer}, \citenamefont
  {Kumar}, \citenamefont {Girvin},\ and\ \citenamefont
  {Schoelkopf}}]{wallraff_strong_2004}%
  \BibitemOpen
  \bibfield  {author} {\bibinfo {author} {\bibfnamefont {A.}~\bibnamefont
  {Wallraff}}, \bibinfo {author} {\bibfnamefont {D.~I.}\ \bibnamefont
  {Schuster}}, \bibinfo {author} {\bibfnamefont {A.}~\bibnamefont {Blais}},
  \bibinfo {author} {\bibfnamefont {L.}~\bibnamefont {Frunzio}}, \bibinfo
  {author} {\bibfnamefont {R.-S.}\ \bibnamefont {Huang}}, \bibinfo {author}
  {\bibfnamefont {J.}~\bibnamefont {Majer}}, \bibinfo {author} {\bibfnamefont
  {S.}~\bibnamefont {Kumar}}, \bibinfo {author} {\bibfnamefont {S.~M.}\
  \bibnamefont {Girvin}}, \ and\ \bibinfo {author} {\bibfnamefont {R.~J.}\
  \bibnamefont {Schoelkopf}},\ }\href {\doibase 10.1038/nature02851} {\bibfield
   {journal} {\bibinfo  {journal} {Nature}\ }\textbf {\bibinfo {volume}
  {431}},\ \bibinfo {pages} {162} (\bibinfo {year} {2004})}\BibitemShut
  {NoStop}%
\bibitem [{\citenamefont {Castellanos-Beltran}\ and\ \citenamefont
  {Lehnert}(2007)}]{castellanos-beltran_widely_2007}%
  \BibitemOpen
  \bibfield  {author} {\bibinfo {author} {\bibfnamefont {M.~A.}\ \bibnamefont
  {Castellanos-Beltran}}\ and\ \bibinfo {author} {\bibfnamefont {K.~W.}\
  \bibnamefont {Lehnert}},\ }\href {\doibase 10.1063/1.2773988} {\bibfield
  {journal} {\bibinfo  {journal} {Applied Physics Letters}\ }\textbf {\bibinfo
  {volume} {91}},\ \bibinfo {pages} {083509} (\bibinfo {year}
  {2007})}\BibitemShut {NoStop}%
\bibitem [{\citenamefont {Vijay}\ \emph {et~al.}(2011)\citenamefont {Vijay},
  \citenamefont {Slichter},\ and\ \citenamefont
  {Siddiqi}}]{vijay_observation_2011}%
  \BibitemOpen
  \bibfield  {author} {\bibinfo {author} {\bibfnamefont {R.}~\bibnamefont
  {Vijay}}, \bibinfo {author} {\bibfnamefont {D.~H.}\ \bibnamefont {Slichter}},
  \ and\ \bibinfo {author} {\bibfnamefont {I.}~\bibnamefont {Siddiqi}},\ }\href
  {\doibase 10.1103/PhysRevLett.106.110502} {\bibfield  {journal} {\bibinfo
  {journal} {Physical Review Letters}\ }\textbf {\bibinfo {volume} {106}}
  (\bibinfo {year} {2011}),\ 10.1103/PhysRevLett.106.110502}\BibitemShut
  {NoStop}%
\bibitem [{\citenamefont {Abdo}\ \emph {et~al.}(2011)\citenamefont {Abdo},
  \citenamefont {Schackert}, \citenamefont {Hatridge}, \citenamefont
  {Rigetti},\ and\ \citenamefont {Devoret}}]{abdo_josephson_2011}%
  \BibitemOpen
  \bibfield  {author} {\bibinfo {author} {\bibfnamefont {B.}~\bibnamefont
  {Abdo}}, \bibinfo {author} {\bibfnamefont {F.}~\bibnamefont {Schackert}},
  \bibinfo {author} {\bibfnamefont {M.}~\bibnamefont {Hatridge}}, \bibinfo
  {author} {\bibfnamefont {C.}~\bibnamefont {Rigetti}}, \ and\ \bibinfo
  {author} {\bibfnamefont {M.}~\bibnamefont {Devoret}},\ }\href {\doibase
  10.1063/1.3653473} {\bibfield  {journal} {\bibinfo  {journal} {Applied
  Physics Letters}\ }\textbf {\bibinfo {volume} {99}},\ \bibinfo {pages}
  {162506} (\bibinfo {year} {2011})}\BibitemShut {NoStop}%
\bibitem [{\citenamefont {Rist{\`e}}\ \emph {et~al.}(2013)\citenamefont
  {Rist{\`e}}, \citenamefont {Dukalski}, \citenamefont {Watson}, \citenamefont
  {{de Lange}}, \citenamefont {Tiggelman}, \citenamefont {Blanter},
  \citenamefont {Lehnert}, \citenamefont {Schouten},\ and\ \citenamefont
  {DiCarlo}}]{riste_deterministic_2013}%
  \BibitemOpen
  \bibfield  {author} {\bibinfo {author} {\bibfnamefont {D.}~\bibnamefont
  {Rist{\`e}}}, \bibinfo {author} {\bibfnamefont {M.}~\bibnamefont {Dukalski}},
  \bibinfo {author} {\bibfnamefont {C.~A.}\ \bibnamefont {Watson}}, \bibinfo
  {author} {\bibfnamefont {G.}~\bibnamefont {{de Lange}}}, \bibinfo {author}
  {\bibfnamefont {M.~J.}\ \bibnamefont {Tiggelman}}, \bibinfo {author}
  {\bibfnamefont {Y.~M.}\ \bibnamefont {Blanter}}, \bibinfo {author}
  {\bibfnamefont {K.~W.}\ \bibnamefont {Lehnert}}, \bibinfo {author}
  {\bibfnamefont {R.~N.}\ \bibnamefont {Schouten}}, \ and\ \bibinfo {author}
  {\bibfnamefont {L.}~\bibnamefont {DiCarlo}},\ }\href {\doibase
  10.1038/nature12513} {\bibfield  {journal} {\bibinfo  {journal} {Nature}\
  }\textbf {\bibinfo {volume} {502}},\ \bibinfo {pages} {350} (\bibinfo {year}
  {2013})}\BibitemShut {NoStop}%
\bibitem [{\citenamefont {Steffen}\ \emph {et~al.}(2013)\citenamefont
  {Steffen}, \citenamefont {Salathe}, \citenamefont {Oppliger}, \citenamefont
  {Kurpiers}, \citenamefont {Baur}, \citenamefont {Lang}, \citenamefont
  {Eichler}, \citenamefont {Puebla-Hellmann}, \citenamefont {Fedorov},\ and\
  \citenamefont {Wallraff}}]{steffen_deterministic_2013}%
  \BibitemOpen
  \bibfield  {author} {\bibinfo {author} {\bibfnamefont {L.}~\bibnamefont
  {Steffen}}, \bibinfo {author} {\bibfnamefont {Y.}~\bibnamefont {Salathe}},
  \bibinfo {author} {\bibfnamefont {M.}~\bibnamefont {Oppliger}}, \bibinfo
  {author} {\bibfnamefont {P.}~\bibnamefont {Kurpiers}}, \bibinfo {author}
  {\bibfnamefont {M.}~\bibnamefont {Baur}}, \bibinfo {author} {\bibfnamefont
  {C.}~\bibnamefont {Lang}}, \bibinfo {author} {\bibfnamefont {C.}~\bibnamefont
  {Eichler}}, \bibinfo {author} {\bibfnamefont {G.}~\bibnamefont
  {Puebla-Hellmann}}, \bibinfo {author} {\bibfnamefont {A.}~\bibnamefont
  {Fedorov}}, \ and\ \bibinfo {author} {\bibfnamefont {A.}~\bibnamefont
  {Wallraff}},\ }\href {\doibase 10.1038/nature12422} {\bibfield  {journal}
  {\bibinfo  {journal} {Nature}\ }\textbf {\bibinfo {volume} {500}},\ \bibinfo
  {pages} {319} (\bibinfo {year} {2013})}\BibitemShut {NoStop}%
\bibitem [{\citenamefont {Chow}\ \emph {et~al.}(2014)\citenamefont {Chow},
  \citenamefont {Gambetta}, \citenamefont {Magesan}, \citenamefont {Abraham},
  \citenamefont {Cross}, \citenamefont {Johnson}, \citenamefont {Masluk},
  \citenamefont {Ryan}, \citenamefont {Smolin}, \citenamefont {Srinivasan},\
  and\ \citenamefont {Steffen}}]{chow_implementing_2014}%
  \BibitemOpen
  \bibfield  {author} {\bibinfo {author} {\bibfnamefont {J.~M.}\ \bibnamefont
  {Chow}}, \bibinfo {author} {\bibfnamefont {J.~M.}\ \bibnamefont {Gambetta}},
  \bibinfo {author} {\bibfnamefont {E.}~\bibnamefont {Magesan}}, \bibinfo
  {author} {\bibfnamefont {D.~W.}\ \bibnamefont {Abraham}}, \bibinfo {author}
  {\bibfnamefont {A.~W.}\ \bibnamefont {Cross}}, \bibinfo {author}
  {\bibfnamefont {B.~R.}\ \bibnamefont {Johnson}}, \bibinfo {author}
  {\bibfnamefont {N.~A.}\ \bibnamefont {Masluk}}, \bibinfo {author}
  {\bibfnamefont {C.~A.}\ \bibnamefont {Ryan}}, \bibinfo {author}
  {\bibfnamefont {J.~A.}\ \bibnamefont {Smolin}}, \bibinfo {author}
  {\bibfnamefont {S.~J.}\ \bibnamefont {Srinivasan}}, \ and\ \bibinfo {author}
  {\bibfnamefont {M.}~\bibnamefont {Steffen}},\ }\href {\doibase
  10.1038/ncomms5015} {\bibfield  {journal} {\bibinfo  {journal} {Nature
  Communications}\ }\textbf {\bibinfo {volume} {5}} (\bibinfo {year} {2014}),\
  10.1038/ncomms5015}\BibitemShut {NoStop}%
\bibitem [{\citenamefont {Roch}\ \emph {et~al.}(2014)\citenamefont {Roch},
  \citenamefont {Schwartz}, \citenamefont {Motzoi}, \citenamefont {Macklin},
  \citenamefont {Vijay}, \citenamefont {Eddins}, \citenamefont {Korotkov},
  \citenamefont {Whaley}, \citenamefont {Sarovar},\ and\ \citenamefont
  {Siddiqi}}]{roch_observation_2014}%
  \BibitemOpen
  \bibfield  {author} {\bibinfo {author} {\bibfnamefont {N.}~\bibnamefont
  {Roch}}, \bibinfo {author} {\bibfnamefont {M.~E.}\ \bibnamefont {Schwartz}},
  \bibinfo {author} {\bibfnamefont {F.}~\bibnamefont {Motzoi}}, \bibinfo
  {author} {\bibfnamefont {C.}~\bibnamefont {Macklin}}, \bibinfo {author}
  {\bibfnamefont {R.}~\bibnamefont {Vijay}}, \bibinfo {author} {\bibfnamefont
  {A.~W.}\ \bibnamefont {Eddins}}, \bibinfo {author} {\bibfnamefont {A.~N.}\
  \bibnamefont {Korotkov}}, \bibinfo {author} {\bibfnamefont {K.~B.}\
  \bibnamefont {Whaley}}, \bibinfo {author} {\bibfnamefont {M.}~\bibnamefont
  {Sarovar}}, \ and\ \bibinfo {author} {\bibfnamefont {I.}~\bibnamefont
  {Siddiqi}},\ }\href {\doibase 10.1103/PhysRevLett.112.170501} {\bibfield
  {journal} {\bibinfo  {journal} {Physical Review Letters}\ }\textbf {\bibinfo
  {volume} {112}} (\bibinfo {year} {2014}),\
  10.1103/PhysRevLett.112.170501}\BibitemShut {NoStop}%
\bibitem [{\citenamefont {Berkley}(2003)}]{berkley_entangled_2003}%
  \BibitemOpen
  \bibfield  {author} {\bibinfo {author} {\bibfnamefont {A.~J.}\ \bibnamefont
  {Berkley}},\ }\href {\doibase 10.1126/science.1084528} {\bibfield  {journal}
  {\bibinfo  {journal} {Science}\ }\textbf {\bibinfo {volume} {300}},\ \bibinfo
  {pages} {1548} (\bibinfo {year} {2003})}\BibitemShut {NoStop}%
\bibitem [{\citenamefont {Steffen}\ \emph {et~al.}(2006)\citenamefont
  {Steffen}, \citenamefont {Ansmann}, \citenamefont {Bialczak}, \citenamefont
  {Katz}, \citenamefont {Lucero}, \citenamefont {McDermott}, \citenamefont
  {Neeley}, \citenamefont {Weig}, \citenamefont {Cleland},\ and\ \citenamefont
  {Martinis}}]{steffen_measurement_2006}%
  \BibitemOpen
  \bibfield  {author} {\bibinfo {author} {\bibfnamefont {M.}~\bibnamefont
  {Steffen}}, \bibinfo {author} {\bibfnamefont {M.}~\bibnamefont {Ansmann}},
  \bibinfo {author} {\bibfnamefont {R.~C.}\ \bibnamefont {Bialczak}}, \bibinfo
  {author} {\bibfnamefont {N.}~\bibnamefont {Katz}}, \bibinfo {author}
  {\bibfnamefont {E.}~\bibnamefont {Lucero}}, \bibinfo {author} {\bibfnamefont
  {R.}~\bibnamefont {McDermott}}, \bibinfo {author} {\bibfnamefont
  {M.}~\bibnamefont {Neeley}}, \bibinfo {author} {\bibfnamefont {E.~M.}\
  \bibnamefont {Weig}}, \bibinfo {author} {\bibfnamefont {A.~N.}\ \bibnamefont
  {Cleland}}, \ and\ \bibinfo {author} {\bibfnamefont {J.~M.}\ \bibnamefont
  {Martinis}},\ }\href {\doibase 10.1126/science.1130886} {\bibfield  {journal}
  {\bibinfo  {journal} {Science}\ }\textbf {\bibinfo {volume} {313}},\ \bibinfo
  {pages} {1423} (\bibinfo {year} {2006})}\BibitemShut {NoStop}%
\bibitem [{\citenamefont {Majer}\ \emph {et~al.}(2007)\citenamefont {Majer},
  \citenamefont {Chow}, \citenamefont {Gambetta}, \citenamefont {Koch},
  \citenamefont {Johnson}, \citenamefont {Schreier}, \citenamefont {Frunzio},
  \citenamefont {Schuster}, \citenamefont {Houck}, \citenamefont {Wallraff},
  \citenamefont {Blais}, \citenamefont {Devoret}, \citenamefont {Girvin},\ and\
  \citenamefont {Schoelkopf}}]{majer_coupling_2007}%
  \BibitemOpen
  \bibfield  {author} {\bibinfo {author} {\bibfnamefont {J.}~\bibnamefont
  {Majer}}, \bibinfo {author} {\bibfnamefont {J.~M.}\ \bibnamefont {Chow}},
  \bibinfo {author} {\bibfnamefont {J.~M.}\ \bibnamefont {Gambetta}}, \bibinfo
  {author} {\bibfnamefont {J.}~\bibnamefont {Koch}}, \bibinfo {author}
  {\bibfnamefont {B.~R.}\ \bibnamefont {Johnson}}, \bibinfo {author}
  {\bibfnamefont {J.~A.}\ \bibnamefont {Schreier}}, \bibinfo {author}
  {\bibfnamefont {L.}~\bibnamefont {Frunzio}}, \bibinfo {author} {\bibfnamefont
  {D.~I.}\ \bibnamefont {Schuster}}, \bibinfo {author} {\bibfnamefont {A.~A.}\
  \bibnamefont {Houck}}, \bibinfo {author} {\bibfnamefont {A.}~\bibnamefont
  {Wallraff}}, \bibinfo {author} {\bibfnamefont {A.}~\bibnamefont {Blais}},
  \bibinfo {author} {\bibfnamefont {M.~H.}\ \bibnamefont {Devoret}}, \bibinfo
  {author} {\bibfnamefont {S.~M.}\ \bibnamefont {Girvin}}, \ and\ \bibinfo
  {author} {\bibfnamefont {R.~J.}\ \bibnamefont {Schoelkopf}},\ }\href
  {\doibase 10.1038/nature06184} {\bibfield  {journal} {\bibinfo  {journal}
  {Nature}\ }\textbf {\bibinfo {volume} {449}},\ \bibinfo {pages} {443}
  (\bibinfo {year} {2007})}\BibitemShut {NoStop}%
\bibitem [{\citenamefont {Shankar}\ \emph {et~al.}(2013)\citenamefont
  {Shankar}, \citenamefont {Hatridge}, \citenamefont {Leghtas}, \citenamefont
  {Sliwa}, \citenamefont {Narla}, \citenamefont {Vool}, \citenamefont {Girvin},
  \citenamefont {Frunzio}, \citenamefont {Mirrahimi},\ and\ \citenamefont
  {Devoret}}]{shankar_2013_autonomously}%
  \BibitemOpen
  \bibfield  {author} {\bibinfo {author} {\bibfnamefont {S.}~\bibnamefont
  {Shankar}}, \bibinfo {author} {\bibfnamefont {M.}~\bibnamefont {Hatridge}},
  \bibinfo {author} {\bibfnamefont {Z.}~\bibnamefont {Leghtas}}, \bibinfo
  {author} {\bibfnamefont {K.}~\bibnamefont {Sliwa}}, \bibinfo {author}
  {\bibfnamefont {A.}~\bibnamefont {Narla}}, \bibinfo {author} {\bibfnamefont
  {U.}~\bibnamefont {Vool}}, \bibinfo {author} {\bibfnamefont {S.~M.}\
  \bibnamefont {Girvin}}, \bibinfo {author} {\bibfnamefont {L.}~\bibnamefont
  {Frunzio}}, \bibinfo {author} {\bibfnamefont {M.}~\bibnamefont {Mirrahimi}},
  \ and\ \bibinfo {author} {\bibfnamefont {M.~H.}\ \bibnamefont {Devoret}},\
  }\href@noop {} {\bibfield  {journal} {\bibinfo  {journal} {Nature}\ }\textbf
  {\bibinfo {volume} {504}},\ \bibinfo {pages} {419} (\bibinfo {year}
  {2013})}\BibitemShut {NoStop}%
\bibitem [{\citenamefont {Geerlings}\ \emph {et~al.}(2013)\citenamefont
  {Geerlings}, \citenamefont {Leghtas}, \citenamefont {Pop}, \citenamefont
  {Shankar}, \citenamefont {Frunzio}, \citenamefont {Schoelkopf}, \citenamefont
  {Mirrahimi},\ and\ \citenamefont {Devoret}}]{geerlings_demonstrating_2013}%
  \BibitemOpen
  \bibfield  {author} {\bibinfo {author} {\bibfnamefont {K.}~\bibnamefont
  {Geerlings}}, \bibinfo {author} {\bibfnamefont {Z.}~\bibnamefont {Leghtas}},
  \bibinfo {author} {\bibfnamefont {I.~M.}\ \bibnamefont {Pop}}, \bibinfo
  {author} {\bibfnamefont {S.}~\bibnamefont {Shankar}}, \bibinfo {author}
  {\bibfnamefont {L.}~\bibnamefont {Frunzio}}, \bibinfo {author} {\bibfnamefont
  {R.~J.}\ \bibnamefont {Schoelkopf}}, \bibinfo {author} {\bibfnamefont
  {M.}~\bibnamefont {Mirrahimi}}, \ and\ \bibinfo {author} {\bibfnamefont
  {M.~H.}\ \bibnamefont {Devoret}},\ }\href {\doibase
  10.1103/PhysRevLett.110.120501} {\bibfield  {journal} {\bibinfo  {journal}
  {Physical Review Letters}\ }\textbf {\bibinfo {volume} {110}} (\bibinfo
  {year} {2013}),\ 10.1103/PhysRevLett.110.120501}\BibitemShut {NoStop}%
\bibitem [{\citenamefont {Minev}\ \emph {et~al.}(2016)\citenamefont {Minev},
  \citenamefont {Serniak}, \citenamefont {Pop}, \citenamefont {Leghtas},
  \citenamefont {Sliwa}, \citenamefont {Hatridge}, \citenamefont {Frunzio},
  \citenamefont {Schoelkopf},\ and\ \citenamefont
  {Devoret}}]{minev_planar_2016}%
  \BibitemOpen
  \bibfield  {author} {\bibinfo {author} {\bibfnamefont {Z.~K.}\ \bibnamefont
  {Minev}}, \bibinfo {author} {\bibfnamefont {K.}~\bibnamefont {Serniak}},
  \bibinfo {author} {\bibfnamefont {I.~M.}\ \bibnamefont {Pop}}, \bibinfo
  {author} {\bibfnamefont {Z.}~\bibnamefont {Leghtas}}, \bibinfo {author}
  {\bibfnamefont {K.}~\bibnamefont {Sliwa}}, \bibinfo {author} {\bibfnamefont
  {M.}~\bibnamefont {Hatridge}}, \bibinfo {author} {\bibfnamefont
  {L.}~\bibnamefont {Frunzio}}, \bibinfo {author} {\bibfnamefont {R.~J.}\
  \bibnamefont {Schoelkopf}}, \ and\ \bibinfo {author} {\bibfnamefont {M.~H.}\
  \bibnamefont {Devoret}},\ }\href {\doibase 10.1103/PhysRevApplied.5.044021}
  {\bibfield  {journal} {\bibinfo  {journal} {Physical Review Applied}\
  }\textbf {\bibinfo {volume} {5}} (\bibinfo {year} {2016}),\
  10.1103/PhysRevApplied.5.044021}\BibitemShut {NoStop}%
\bibitem [{\citenamefont {Brecht}\ \emph {et~al.}(2015)\citenamefont {Brecht},
  \citenamefont {Reagor}, \citenamefont {Chu}, \citenamefont {Pfaff},
  \citenamefont {Wang}, \citenamefont {Frunzio}, \citenamefont {Devoret},\ and\
  \citenamefont {Schoelkopf}}]{brecht_demonstration_2015}%
  \BibitemOpen
  \bibfield  {author} {\bibinfo {author} {\bibfnamefont {T.}~\bibnamefont
  {Brecht}}, \bibinfo {author} {\bibfnamefont {M.}~\bibnamefont {Reagor}},
  \bibinfo {author} {\bibfnamefont {Y.}~\bibnamefont {Chu}}, \bibinfo {author}
  {\bibfnamefont {W.}~\bibnamefont {Pfaff}}, \bibinfo {author} {\bibfnamefont
  {C.}~\bibnamefont {Wang}}, \bibinfo {author} {\bibfnamefont {L.}~\bibnamefont
  {Frunzio}}, \bibinfo {author} {\bibfnamefont {M.~H.}\ \bibnamefont
  {Devoret}}, \ and\ \bibinfo {author} {\bibfnamefont {R.~J.}\ \bibnamefont
  {Schoelkopf}},\ }\href {\doibase 10.1063/1.4935541} {\bibfield  {journal}
  {\bibinfo  {journal} {Applied Physics Letters}\ }\textbf {\bibinfo {volume}
  {107}},\ \bibinfo {pages} {192603} (\bibinfo {year} {2015})}\BibitemShut
  {NoStop}%
\bibitem [{\citenamefont {Kelly}\ \emph {et~al.}(2015)\citenamefont {Kelly},
  \citenamefont {Barends}, \citenamefont {Fowler}, \citenamefont {Megrant},
  \citenamefont {Jeffrey}, \citenamefont {White}, \citenamefont {Sank},
  \citenamefont {Mutus}, \citenamefont {Campbell}, \citenamefont {Chen},
  \citenamefont {Chen}, \citenamefont {Chiaro}, \citenamefont {Dunsworth},
  \citenamefont {Hoi}, \citenamefont {Neill}, \citenamefont {O'Malley},
  \citenamefont {Quintana}, \citenamefont {Roushan}, \citenamefont
  {Vainsencher}, \citenamefont {Wenner}, \citenamefont {Cleland},\ and\
  \citenamefont {Martinis}}]{kelly_state_2015}%
  \BibitemOpen
  \bibfield  {author} {\bibinfo {author} {\bibfnamefont {J.}~\bibnamefont
  {Kelly}}, \bibinfo {author} {\bibfnamefont {R.}~\bibnamefont {Barends}},
  \bibinfo {author} {\bibfnamefont {A.~G.}\ \bibnamefont {Fowler}}, \bibinfo
  {author} {\bibfnamefont {A.}~\bibnamefont {Megrant}}, \bibinfo {author}
  {\bibfnamefont {E.}~\bibnamefont {Jeffrey}}, \bibinfo {author} {\bibfnamefont
  {T.~C.}\ \bibnamefont {White}}, \bibinfo {author} {\bibfnamefont
  {D.}~\bibnamefont {Sank}}, \bibinfo {author} {\bibfnamefont {J.~Y.}\
  \bibnamefont {Mutus}}, \bibinfo {author} {\bibfnamefont {B.}~\bibnamefont
  {Campbell}}, \bibinfo {author} {\bibfnamefont {Y.}~\bibnamefont {Chen}},
  \bibinfo {author} {\bibfnamefont {Z.}~\bibnamefont {Chen}}, \bibinfo {author}
  {\bibfnamefont {B.}~\bibnamefont {Chiaro}}, \bibinfo {author} {\bibfnamefont
  {A.}~\bibnamefont {Dunsworth}}, \bibinfo {author} {\bibfnamefont {I.-C.}\
  \bibnamefont {Hoi}}, \bibinfo {author} {\bibfnamefont {C.}~\bibnamefont
  {Neill}}, \bibinfo {author} {\bibfnamefont {P.~J.~J.}\ \bibnamefont
  {O'Malley}}, \bibinfo {author} {\bibfnamefont {C.}~\bibnamefont {Quintana}},
  \bibinfo {author} {\bibfnamefont {P.}~\bibnamefont {Roushan}}, \bibinfo
  {author} {\bibfnamefont {A.}~\bibnamefont {Vainsencher}}, \bibinfo {author}
  {\bibfnamefont {J.}~\bibnamefont {Wenner}}, \bibinfo {author} {\bibfnamefont
  {A.~N.}\ \bibnamefont {Cleland}}, \ and\ \bibinfo {author} {\bibfnamefont
  {J.~M.}\ \bibnamefont {Martinis}},\ }\href {\doibase 10.1038/nature14270}
  {\bibfield  {journal} {\bibinfo  {journal} {Nature}\ }\textbf {\bibinfo
  {volume} {519}},\ \bibinfo {pages} {66} (\bibinfo {year} {2015})}\BibitemShut
  {NoStop}%
\bibitem [{\citenamefont {Rist{\`e}}\ \emph {et~al.}(2015)\citenamefont
  {Rist{\`e}}, \citenamefont {Poletto}, \citenamefont {Huang}, \citenamefont
  {Bruno}, \citenamefont {Vesterinen}, \citenamefont {Saira},\ and\
  \citenamefont {DiCarlo}}]{riste_detecting_2015}%
  \BibitemOpen
  \bibfield  {author} {\bibinfo {author} {\bibfnamefont {D.}~\bibnamefont
  {Rist{\`e}}}, \bibinfo {author} {\bibfnamefont {S.}~\bibnamefont {Poletto}},
  \bibinfo {author} {\bibfnamefont {M.-Z.}\ \bibnamefont {Huang}}, \bibinfo
  {author} {\bibfnamefont {A.}~\bibnamefont {Bruno}}, \bibinfo {author}
  {\bibfnamefont {V.}~\bibnamefont {Vesterinen}}, \bibinfo {author}
  {\bibfnamefont {O.-P.}\ \bibnamefont {Saira}}, \ and\ \bibinfo {author}
  {\bibfnamefont {L.}~\bibnamefont {DiCarlo}},\ }\href {\doibase
  10.1038/ncomms7983} {\bibfield  {journal} {\bibinfo  {journal} {Nature
  Communications}\ }\textbf {\bibinfo {volume} {6}},\ \bibinfo {pages} {6983}
  (\bibinfo {year} {2015})}\BibitemShut {NoStop}%
\bibitem [{\citenamefont {Paik}\ \emph {et~al.}(2011)\citenamefont {Paik},
  \citenamefont {Schuster}, \citenamefont {Bishop}, \citenamefont {Kirchmair},
  \citenamefont {Catelani}, \citenamefont {Sears}, \citenamefont {Johnson},
  \citenamefont {Reagor}, \citenamefont {Frunzio}, \citenamefont {Glazman},
  \citenamefont {Girvin}, \citenamefont {Devoret},\ and\ \citenamefont
  {Schoelkopf}}]{paik_observation_2011}%
  \BibitemOpen
  \bibfield  {author} {\bibinfo {author} {\bibfnamefont {H.}~\bibnamefont
  {Paik}}, \bibinfo {author} {\bibfnamefont {D.~I.}\ \bibnamefont {Schuster}},
  \bibinfo {author} {\bibfnamefont {L.~S.}\ \bibnamefont {Bishop}}, \bibinfo
  {author} {\bibfnamefont {G.}~\bibnamefont {Kirchmair}}, \bibinfo {author}
  {\bibfnamefont {G.}~\bibnamefont {Catelani}}, \bibinfo {author}
  {\bibfnamefont {A.~P.}\ \bibnamefont {Sears}}, \bibinfo {author}
  {\bibfnamefont {B.~R.}\ \bibnamefont {Johnson}}, \bibinfo {author}
  {\bibfnamefont {M.~J.}\ \bibnamefont {Reagor}}, \bibinfo {author}
  {\bibfnamefont {L.}~\bibnamefont {Frunzio}}, \bibinfo {author} {\bibfnamefont
  {L.~I.}\ \bibnamefont {Glazman}}, \bibinfo {author} {\bibfnamefont {S.~M.}\
  \bibnamefont {Girvin}}, \bibinfo {author} {\bibfnamefont {M.~H.}\
  \bibnamefont {Devoret}}, \ and\ \bibinfo {author} {\bibfnamefont {R.~J.}\
  \bibnamefont {Schoelkopf}},\ }\href {\doibase 10.1103/PhysRevLett.107.240501}
  {\bibfield  {journal} {\bibinfo  {journal} {Physical Review Letters}\
  }\textbf {\bibinfo {volume} {107}} (\bibinfo {year} {2011}),\
  10.1103/PhysRevLett.107.240501}\BibitemShut {NoStop}%
\bibitem [{\citenamefont {Rigetti}\ \emph {et~al.}(2012)\citenamefont
  {Rigetti}, \citenamefont {Gambetta}, \citenamefont {Poletto}, \citenamefont
  {Plourde}, \citenamefont {Chow}, \citenamefont {C{\'o}rcoles}, \citenamefont
  {Smolin}, \citenamefont {Merkel}, \citenamefont {Rozen}, \citenamefont
  {Keefe}, \citenamefont {Rothwell}, \citenamefont {Ketchen},\ and\
  \citenamefont {Steffen}}]{rigetti_superconducting_2012}%
  \BibitemOpen
  \bibfield  {author} {\bibinfo {author} {\bibfnamefont {C.}~\bibnamefont
  {Rigetti}}, \bibinfo {author} {\bibfnamefont {J.~M.}\ \bibnamefont
  {Gambetta}}, \bibinfo {author} {\bibfnamefont {S.}~\bibnamefont {Poletto}},
  \bibinfo {author} {\bibfnamefont {B.~L.~T.}\ \bibnamefont {Plourde}},
  \bibinfo {author} {\bibfnamefont {J.~M.}\ \bibnamefont {Chow}}, \bibinfo
  {author} {\bibfnamefont {A.~D.}\ \bibnamefont {C{\'o}rcoles}}, \bibinfo
  {author} {\bibfnamefont {J.~A.}\ \bibnamefont {Smolin}}, \bibinfo {author}
  {\bibfnamefont {S.~T.}\ \bibnamefont {Merkel}}, \bibinfo {author}
  {\bibfnamefont {J.~R.}\ \bibnamefont {Rozen}}, \bibinfo {author}
  {\bibfnamefont {G.~A.}\ \bibnamefont {Keefe}}, \bibinfo {author}
  {\bibfnamefont {M.~B.}\ \bibnamefont {Rothwell}}, \bibinfo {author}
  {\bibfnamefont {M.~B.}\ \bibnamefont {Ketchen}}, \ and\ \bibinfo {author}
  {\bibfnamefont {M.}~\bibnamefont {Steffen}},\ }\href {\doibase
  10.1103/PhysRevB.86.100506} {\bibfield  {journal} {\bibinfo  {journal}
  {Physical Review B}\ }\textbf {\bibinfo {volume} {86}} (\bibinfo {year}
  {2012}),\ 10.1103/PhysRevB.86.100506}\BibitemShut {NoStop}%
\bibitem [{\citenamefont {Barends}\ \emph {et~al.}(2013)\citenamefont
  {Barends}, \citenamefont {Kelly}, \citenamefont {Megrant}, \citenamefont
  {Sank}, \citenamefont {Jeffrey}, \citenamefont {Chen}, \citenamefont {Yin},
  \citenamefont {Chiaro}, \citenamefont {Mutus}, \citenamefont {Neill},
  \citenamefont {O'Malley}, \citenamefont {Roushan}, \citenamefont {Wenner},
  \citenamefont {White}, \citenamefont {Cleland},\ and\ \citenamefont
  {Martinis}}]{barends_coherent_2013}%
  \BibitemOpen
  \bibfield  {author} {\bibinfo {author} {\bibfnamefont {R.}~\bibnamefont
  {Barends}}, \bibinfo {author} {\bibfnamefont {J.}~\bibnamefont {Kelly}},
  \bibinfo {author} {\bibfnamefont {A.}~\bibnamefont {Megrant}}, \bibinfo
  {author} {\bibfnamefont {D.}~\bibnamefont {Sank}}, \bibinfo {author}
  {\bibfnamefont {E.}~\bibnamefont {Jeffrey}}, \bibinfo {author} {\bibfnamefont
  {Y.}~\bibnamefont {Chen}}, \bibinfo {author} {\bibfnamefont {Y.}~\bibnamefont
  {Yin}}, \bibinfo {author} {\bibfnamefont {B.}~\bibnamefont {Chiaro}},
  \bibinfo {author} {\bibfnamefont {J.}~\bibnamefont {Mutus}}, \bibinfo
  {author} {\bibfnamefont {C.}~\bibnamefont {Neill}}, \bibinfo {author}
  {\bibfnamefont {P.}~\bibnamefont {O'Malley}}, \bibinfo {author}
  {\bibfnamefont {P.}~\bibnamefont {Roushan}}, \bibinfo {author} {\bibfnamefont
  {J.}~\bibnamefont {Wenner}}, \bibinfo {author} {\bibfnamefont {T.~C.}\
  \bibnamefont {White}}, \bibinfo {author} {\bibfnamefont {A.~N.}\ \bibnamefont
  {Cleland}}, \ and\ \bibinfo {author} {\bibfnamefont {J.~M.}\ \bibnamefont
  {Martinis}},\ }\href {\doibase 10.1103/PhysRevLett.111.080502} {\bibfield
  {journal} {\bibinfo  {journal} {Physical Review Letters}\ }\textbf {\bibinfo
  {volume} {111}} (\bibinfo {year} {2013}),\
  10.1103/PhysRevLett.111.080502}\BibitemShut {NoStop}%
\bibitem [{\citenamefont {Sandberg}\ \emph {et~al.}(2012)\citenamefont
  {Sandberg}, \citenamefont {Vissers}, \citenamefont {Kline}, \citenamefont
  {Weides}, \citenamefont {Gao}, \citenamefont {Wisbey},\ and\ \citenamefont
  {Pappas}}]{sandberg_etch_2012}%
  \BibitemOpen
  \bibfield  {author} {\bibinfo {author} {\bibfnamefont {M.}~\bibnamefont
  {Sandberg}}, \bibinfo {author} {\bibfnamefont {M.~R.}\ \bibnamefont
  {Vissers}}, \bibinfo {author} {\bibfnamefont {J.~S.}\ \bibnamefont {Kline}},
  \bibinfo {author} {\bibfnamefont {M.}~\bibnamefont {Weides}}, \bibinfo
  {author} {\bibfnamefont {J.}~\bibnamefont {Gao}}, \bibinfo {author}
  {\bibfnamefont {D.~S.}\ \bibnamefont {Wisbey}}, \ and\ \bibinfo {author}
  {\bibfnamefont {D.~P.}\ \bibnamefont {Pappas}},\ }\href {\doibase
  10.1063/1.4729623} {\bibfield  {journal} {\bibinfo  {journal} {Applied
  Physics Letters}\ }\textbf {\bibinfo {volume} {100}},\ \bibinfo {pages}
  {262605} (\bibinfo {year} {2012})}\BibitemShut {NoStop}%
\bibitem [{\citenamefont {Vissers}\ \emph {et~al.}(2012)\citenamefont
  {Vissers}, \citenamefont {Weides}, \citenamefont {Kline}, \citenamefont
  {Sandberg},\ and\ \citenamefont {Pappas}}]{vissers_identifying_2012}%
  \BibitemOpen
  \bibfield  {author} {\bibinfo {author} {\bibfnamefont {M.~R.}\ \bibnamefont
  {Vissers}}, \bibinfo {author} {\bibfnamefont {M.~P.}\ \bibnamefont {Weides}},
  \bibinfo {author} {\bibfnamefont {J.~S.}\ \bibnamefont {Kline}}, \bibinfo
  {author} {\bibfnamefont {M.}~\bibnamefont {Sandberg}}, \ and\ \bibinfo
  {author} {\bibfnamefont {D.~P.}\ \bibnamefont {Pappas}},\ }\href {\doibase
  10.1063/1.4730389} {\bibfield  {journal} {\bibinfo  {journal} {Applied
  Physics Letters}\ }\textbf {\bibinfo {volume} {101}},\ \bibinfo {pages}
  {022601} (\bibinfo {year} {2012})}\BibitemShut {NoStop}%
\bibitem [{\citenamefont {Braum{\"u}ller}\ \emph {et~al.}(2015)\citenamefont
  {Braum{\"u}ller}, \citenamefont {Cramer}, \citenamefont {Schl{\"o}r},
  \citenamefont {Rotzinger}, \citenamefont {Radtke}, \citenamefont
  {Lukashenko}, \citenamefont {Yang}, \citenamefont {Skacel}, \citenamefont
  {Probst}, \citenamefont {Marthaler}, \citenamefont {Guo}, \citenamefont
  {Ustinov},\ and\ \citenamefont {Weides}}]{braumuller_multiphoton_2015}%
  \BibitemOpen
  \bibfield  {author} {\bibinfo {author} {\bibfnamefont {J.}~\bibnamefont
  {Braum{\"u}ller}}, \bibinfo {author} {\bibfnamefont {J.}~\bibnamefont
  {Cramer}}, \bibinfo {author} {\bibfnamefont {S.}~\bibnamefont {Schl{\"o}r}},
  \bibinfo {author} {\bibfnamefont {H.}~\bibnamefont {Rotzinger}}, \bibinfo
  {author} {\bibfnamefont {L.}~\bibnamefont {Radtke}}, \bibinfo {author}
  {\bibfnamefont {A.}~\bibnamefont {Lukashenko}}, \bibinfo {author}
  {\bibfnamefont {P.}~\bibnamefont {Yang}}, \bibinfo {author} {\bibfnamefont
  {S.~T.}\ \bibnamefont {Skacel}}, \bibinfo {author} {\bibfnamefont
  {S.}~\bibnamefont {Probst}}, \bibinfo {author} {\bibfnamefont
  {M.}~\bibnamefont {Marthaler}}, \bibinfo {author} {\bibfnamefont
  {L.}~\bibnamefont {Guo}}, \bibinfo {author} {\bibfnamefont {A.~V.}\
  \bibnamefont {Ustinov}}, \ and\ \bibinfo {author} {\bibfnamefont
  {M.}~\bibnamefont {Weides}},\ }\href {\doibase 10.1103/PhysRevB.91.054523}
  {\bibfield  {journal} {\bibinfo  {journal} {Physical Review B}\ }\textbf
  {\bibinfo {volume} {91}} (\bibinfo {year} {2015}),\
  10.1103/PhysRevB.91.054523}\BibitemShut {NoStop}%
\bibitem [{\citenamefont {O'Connell}\ \emph {et~al.}(2008)\citenamefont
  {O'Connell}, \citenamefont {Ansmann}, \citenamefont {Bialczak}, \citenamefont
  {Hofheinz}, \citenamefont {Katz}, \citenamefont {Lucero}, \citenamefont
  {McKenney}, \citenamefont {Neeley}, \citenamefont {Wang}, \citenamefont
  {Weig}, \citenamefont {Cleland},\ and\ \citenamefont
  {Martinis}}]{oconnell_microwave_2008}%
  \BibitemOpen
  \bibfield  {author} {\bibinfo {author} {\bibfnamefont {A.~D.}\ \bibnamefont
  {O'Connell}}, \bibinfo {author} {\bibfnamefont {M.}~\bibnamefont {Ansmann}},
  \bibinfo {author} {\bibfnamefont {R.~C.}\ \bibnamefont {Bialczak}}, \bibinfo
  {author} {\bibfnamefont {M.}~\bibnamefont {Hofheinz}}, \bibinfo {author}
  {\bibfnamefont {N.}~\bibnamefont {Katz}}, \bibinfo {author} {\bibfnamefont
  {E.}~\bibnamefont {Lucero}}, \bibinfo {author} {\bibfnamefont
  {C.}~\bibnamefont {McKenney}}, \bibinfo {author} {\bibfnamefont
  {M.}~\bibnamefont {Neeley}}, \bibinfo {author} {\bibfnamefont
  {H.}~\bibnamefont {Wang}}, \bibinfo {author} {\bibfnamefont {E.~M.}\
  \bibnamefont {Weig}}, \bibinfo {author} {\bibfnamefont {A.~N.}\ \bibnamefont
  {Cleland}}, \ and\ \bibinfo {author} {\bibfnamefont {J.~M.}\ \bibnamefont
  {Martinis}},\ }\href {\doibase 10.1063/1.2898887} {\bibfield  {journal}
  {\bibinfo  {journal} {Applied Physics Letters}\ }\textbf {\bibinfo {volume}
  {92}},\ \bibinfo {pages} {112903} (\bibinfo {year} {2008})}\BibitemShut
  {NoStop}%
\bibitem [{\citenamefont {Wang}\ \emph {et~al.}(2015)\citenamefont {Wang},
  \citenamefont {Axline}, \citenamefont {Gao}, \citenamefont {Brecht},
  \citenamefont {Chu}, \citenamefont {Frunzio}, \citenamefont {Devoret},\ and\
  \citenamefont {Schoelkopf}}]{wang_surface_2015}%
  \BibitemOpen
  \bibfield  {author} {\bibinfo {author} {\bibfnamefont {C.}~\bibnamefont
  {Wang}}, \bibinfo {author} {\bibfnamefont {C.}~\bibnamefont {Axline}},
  \bibinfo {author} {\bibfnamefont {Y.~Y.}\ \bibnamefont {Gao}}, \bibinfo
  {author} {\bibfnamefont {T.}~\bibnamefont {Brecht}}, \bibinfo {author}
  {\bibfnamefont {Y.}~\bibnamefont {Chu}}, \bibinfo {author} {\bibfnamefont
  {L.}~\bibnamefont {Frunzio}}, \bibinfo {author} {\bibfnamefont {M.~H.}\
  \bibnamefont {Devoret}}, \ and\ \bibinfo {author} {\bibfnamefont {R.~J.}\
  \bibnamefont {Schoelkopf}},\ }\href {\doibase 10.1063/1.4934486} {\bibfield
  {journal} {\bibinfo  {journal} {Applied Physics Letters}\ }\textbf {\bibinfo
  {volume} {107}},\ \bibinfo {pages} {162601} (\bibinfo {year}
  {2015})}\BibitemShut {NoStop}%
\bibitem [{\citenamefont {Kou}\ \emph {et~al.}(2017)\citenamefont {Kou},
  \citenamefont {Smith}, \citenamefont {Vool}, \citenamefont {Pop},
  \citenamefont {Sliwa}, \citenamefont {Hatridge}, \citenamefont {Frunzio},\
  and\ \citenamefont {Devoret}}]{kou_simultaneous_2017}%
  \BibitemOpen
  \bibfield  {author} {\bibinfo {author} {\bibfnamefont {A.}~\bibnamefont
  {Kou}}, \bibinfo {author} {\bibfnamefont {W.~C.}\ \bibnamefont {Smith}},
  \bibinfo {author} {\bibfnamefont {U.}~\bibnamefont {Vool}}, \bibinfo {author}
  {\bibfnamefont {I.~M.}\ \bibnamefont {Pop}}, \bibinfo {author} {\bibfnamefont
  {K.~M.}\ \bibnamefont {Sliwa}}, \bibinfo {author} {\bibfnamefont {M.~H.}\
  \bibnamefont {Hatridge}}, \bibinfo {author} {\bibfnamefont {L.}~\bibnamefont
  {Frunzio}}, \ and\ \bibinfo {author} {\bibfnamefont {M.~H.}\ \bibnamefont
  {Devoret}},\ }\href@noop {} {\bibfield  {journal} {\bibinfo  {journal}
  {arXiv:1705.05712 [quant-ph]}\ } (\bibinfo {year} {2017})}\BibitemShut
  {NoStop}%
\bibitem [{\citenamefont {Barends}\ \emph {et~al.}(2011)\citenamefont
  {Barends}, \citenamefont {Wenner}, \citenamefont {Lenander}, \citenamefont
  {Chen}, \citenamefont {Bialczak}, \citenamefont {Kelly}, \citenamefont
  {Lucero}, \citenamefont {O'Malley}, \citenamefont {Mariantoni}, \citenamefont
  {Sank}, \citenamefont {Wang}, \citenamefont {White}, \citenamefont {Yin},
  \citenamefont {Zhao}, \citenamefont {Cleland}, \citenamefont {Martinis},\
  and\ \citenamefont {Baselmans}}]{barends_minimizing_2011}%
  \BibitemOpen
  \bibfield  {author} {\bibinfo {author} {\bibfnamefont {R.}~\bibnamefont
  {Barends}}, \bibinfo {author} {\bibfnamefont {J.}~\bibnamefont {Wenner}},
  \bibinfo {author} {\bibfnamefont {M.}~\bibnamefont {Lenander}}, \bibinfo
  {author} {\bibfnamefont {Y.}~\bibnamefont {Chen}}, \bibinfo {author}
  {\bibfnamefont {R.~C.}\ \bibnamefont {Bialczak}}, \bibinfo {author}
  {\bibfnamefont {J.}~\bibnamefont {Kelly}}, \bibinfo {author} {\bibfnamefont
  {E.}~\bibnamefont {Lucero}}, \bibinfo {author} {\bibfnamefont
  {P.}~\bibnamefont {O'Malley}}, \bibinfo {author} {\bibfnamefont
  {M.}~\bibnamefont {Mariantoni}}, \bibinfo {author} {\bibfnamefont
  {D.}~\bibnamefont {Sank}}, \bibinfo {author} {\bibfnamefont {H.}~\bibnamefont
  {Wang}}, \bibinfo {author} {\bibfnamefont {T.~C.}\ \bibnamefont {White}},
  \bibinfo {author} {\bibfnamefont {Y.}~\bibnamefont {Yin}}, \bibinfo {author}
  {\bibfnamefont {J.}~\bibnamefont {Zhao}}, \bibinfo {author} {\bibfnamefont
  {A.~N.}\ \bibnamefont {Cleland}}, \bibinfo {author} {\bibfnamefont {J.~M.}\
  \bibnamefont {Martinis}}, \ and\ \bibinfo {author} {\bibfnamefont {J.~J.~A.}\
  \bibnamefont {Baselmans}},\ }\href {\doibase 10.1063/1.3638063} {\bibfield
  {journal} {\bibinfo  {journal} {Applied Physics Letters}\ }\textbf {\bibinfo
  {volume} {99}},\ \bibinfo {pages} {113507} (\bibinfo {year}
  {2011})}\BibitemShut {NoStop}%
\bibitem [{\citenamefont {Probst}\ \emph {et~al.}(2015)\citenamefont {Probst},
  \citenamefont {Song}, \citenamefont {Bushev}, \citenamefont {Ustinov},\ and\
  \citenamefont {Weides}}]{probst_efficient_2015}%
  \BibitemOpen
  \bibfield  {author} {\bibinfo {author} {\bibfnamefont {S.}~\bibnamefont
  {Probst}}, \bibinfo {author} {\bibfnamefont {F.~B.}\ \bibnamefont {Song}},
  \bibinfo {author} {\bibfnamefont {P.~A.}\ \bibnamefont {Bushev}}, \bibinfo
  {author} {\bibfnamefont {A.~V.}\ \bibnamefont {Ustinov}}, \ and\ \bibinfo
  {author} {\bibfnamefont {M.}~\bibnamefont {Weides}},\ }\href {\doibase
  10.1063/1.4907935} {\bibfield  {journal} {\bibinfo  {journal} {Review of
  Scientific Instruments}\ }\textbf {\bibinfo {volume} {86}},\ \bibinfo {pages}
  {024706} (\bibinfo {year} {2015})}\BibitemShut {NoStop}%
\bibitem [{\citenamefont {Gao}(2008)}]{gao_physics_2008}%
  \BibitemOpen
  \bibfield  {author} {\bibinfo {author} {\bibfnamefont {J.}~\bibnamefont
  {Gao}},\ }\emph {\bibinfo {title} {The Physics of Superconducting Microwave
  Resonators}},\ \href@noop {} {\bibinfo {type} {Phd dissertation}},\ \bibinfo
  {school} {California Institute of Technology} (\bibinfo {year}
  {2008})\BibitemShut {NoStop}%
\bibitem [{\citenamefont {Dunsworth}\ \emph {et~al.}(2017)\citenamefont
  {Dunsworth}, \citenamefont {Megrant}, \citenamefont {Quintana}, \citenamefont
  {Chen}, \citenamefont {Barends}, \citenamefont {Burkett}, \citenamefont
  {Foxen}, \citenamefont {Chen}, \citenamefont {Chiaro}, \citenamefont
  {Fowler}, \citenamefont {Graff}, \citenamefont {Jeffrey}, \citenamefont
  {Kelly}, \citenamefont {Lucero}, \citenamefont {Mutus}, \citenamefont
  {Neeley}, \citenamefont {Neill}, \citenamefont {Roushan}, \citenamefont
  {Sank}, \citenamefont {Vainsencher}, \citenamefont {Wenner}, \citenamefont
  {White},\ and\ \citenamefont {Martinis}}]{dunsworth_characterization_2017}%
  \BibitemOpen
  \bibfield  {author} {\bibinfo {author} {\bibfnamefont {A.}~\bibnamefont
  {Dunsworth}}, \bibinfo {author} {\bibfnamefont {A.}~\bibnamefont {Megrant}},
  \bibinfo {author} {\bibfnamefont {C.}~\bibnamefont {Quintana}}, \bibinfo
  {author} {\bibfnamefont {Z.}~\bibnamefont {Chen}}, \bibinfo {author}
  {\bibfnamefont {R.}~\bibnamefont {Barends}}, \bibinfo {author} {\bibfnamefont
  {B.}~\bibnamefont {Burkett}}, \bibinfo {author} {\bibfnamefont
  {B.}~\bibnamefont {Foxen}}, \bibinfo {author} {\bibfnamefont
  {Y.}~\bibnamefont {Chen}}, \bibinfo {author} {\bibfnamefont {B.}~\bibnamefont
  {Chiaro}}, \bibinfo {author} {\bibfnamefont {A.}~\bibnamefont {Fowler}},
  \bibinfo {author} {\bibfnamefont {R.}~\bibnamefont {Graff}}, \bibinfo
  {author} {\bibfnamefont {E.}~\bibnamefont {Jeffrey}}, \bibinfo {author}
  {\bibfnamefont {J.}~\bibnamefont {Kelly}}, \bibinfo {author} {\bibfnamefont
  {E.}~\bibnamefont {Lucero}}, \bibinfo {author} {\bibfnamefont
  {J.}~\bibnamefont {Mutus}}, \bibinfo {author} {\bibfnamefont
  {M.}~\bibnamefont {Neeley}}, \bibinfo {author} {\bibfnamefont
  {C.}~\bibnamefont {Neill}}, \bibinfo {author} {\bibfnamefont
  {P.}~\bibnamefont {Roushan}}, \bibinfo {author} {\bibfnamefont
  {D.}~\bibnamefont {Sank}}, \bibinfo {author} {\bibfnamefont {A.}~\bibnamefont
  {Vainsencher}}, \bibinfo {author} {\bibfnamefont {J.}~\bibnamefont {Wenner}},
  \bibinfo {author} {\bibfnamefont {T.}~\bibnamefont {White}}, \ and\ \bibinfo
  {author} {\bibfnamefont {J.~M.}\ \bibnamefont {Martinis}},\ }\href
  {https://arxiv.org/abs/1706.00879} {\bibfield  {journal} {\bibinfo  {journal}
  {arXiv:1706.00879 [quant-ph]}\ } (\bibinfo {year} {2017})}\BibitemShut
  {NoStop}%
\bibitem [{\citenamefont {Wu}\ \emph {et~al.}(2017)\citenamefont {Wu},
  \citenamefont {Long}, \citenamefont {Ku}, \citenamefont {Lake}, \citenamefont
  {Bal},\ and\ \citenamefont {Pappas}}]{wu_overlap_2017}%
  \BibitemOpen
  \bibfield  {author} {\bibinfo {author} {\bibfnamefont {X.}~\bibnamefont
  {Wu}}, \bibinfo {author} {\bibfnamefont {J.~L.}\ \bibnamefont {Long}},
  \bibinfo {author} {\bibfnamefont {H.~S.}\ \bibnamefont {Ku}}, \bibinfo
  {author} {\bibfnamefont {R.~E.}\ \bibnamefont {Lake}}, \bibinfo {author}
  {\bibfnamefont {M.}~\bibnamefont {Bal}}, \ and\ \bibinfo {author}
  {\bibfnamefont {D.~P.}\ \bibnamefont {Pappas}},\ }\href
  {https://arxiv.org/abs/1705.08993} {\bibfield  {journal} {\bibinfo  {journal}
  {arXiv:1705.08993 [cond-mat.supr-con]}\ } (\bibinfo {year}
  {2017})}\BibitemShut {NoStop}%
\bibitem [{\citenamefont {Mooij}\ and\ \citenamefont
  {Nazarov}(2006)}]{mooij_superconducting_2006}%
  \BibitemOpen
  \bibfield  {author} {\bibinfo {author} {\bibfnamefont {J.~E.}\ \bibnamefont
  {Mooij}}\ and\ \bibinfo {author} {\bibfnamefont {Y.~V.}\ \bibnamefont
  {Nazarov}},\ }\href {\doibase 10.1038/nphys234} {\bibfield  {journal}
  {\bibinfo  {journal} {Nature Physics}\ }\textbf {\bibinfo {volume} {2}},\
  \bibinfo {pages} {169} (\bibinfo {year} {2006})}\BibitemShut {NoStop}%
\bibitem [{\citenamefont {Astafiev}\ \emph {et~al.}(2012)\citenamefont
  {Astafiev}, \citenamefont {Ioffe}, \citenamefont {Kafanov}, \citenamefont
  {Pashkin}, \citenamefont {Arutyunov}, \citenamefont {Shahar}, \citenamefont
  {Cohen},\ and\ \citenamefont {Tsai}}]{astafiev_coherent_2012}%
  \BibitemOpen
  \bibfield  {author} {\bibinfo {author} {\bibfnamefont {O.~V.}\ \bibnamefont
  {Astafiev}}, \bibinfo {author} {\bibfnamefont {L.~B.}\ \bibnamefont {Ioffe}},
  \bibinfo {author} {\bibfnamefont {S.}~\bibnamefont {Kafanov}}, \bibinfo
  {author} {\bibfnamefont {Y.~A.}\ \bibnamefont {Pashkin}}, \bibinfo {author}
  {\bibfnamefont {K.~Y.}\ \bibnamefont {Arutyunov}}, \bibinfo {author}
  {\bibfnamefont {D.}~\bibnamefont {Shahar}}, \bibinfo {author} {\bibfnamefont
  {O.}~\bibnamefont {Cohen}}, \ and\ \bibinfo {author} {\bibfnamefont {J.~S.}\
  \bibnamefont {Tsai}},\ }\href {\doibase 10.1038/nature10930} {\bibfield
  {journal} {\bibinfo  {journal} {Nature}\ }\textbf {\bibinfo {volume} {484}},\
  \bibinfo {pages} {355} (\bibinfo {year} {2012})}\BibitemShut {NoStop}%
\bibitem [{\citenamefont {Cleuziou}\ \emph {et~al.}(2006)\citenamefont
  {Cleuziou}, \citenamefont {Wernsdorfer}, \citenamefont {Bouchiat},
  \citenamefont {Ondar{\c c}uhu},\ and\ \citenamefont
  {Monthioux}}]{cleuziou_carbon_2006}%
  \BibitemOpen
  \bibfield  {author} {\bibinfo {author} {\bibfnamefont {J.-P.}\ \bibnamefont
  {Cleuziou}}, \bibinfo {author} {\bibfnamefont {W.}~\bibnamefont
  {Wernsdorfer}}, \bibinfo {author} {\bibfnamefont {V.}~\bibnamefont
  {Bouchiat}}, \bibinfo {author} {\bibfnamefont {T.}~\bibnamefont {Ondar{\c
  c}uhu}}, \ and\ \bibinfo {author} {\bibfnamefont {M.}~\bibnamefont
  {Monthioux}},\ }\href {\doibase 10.1038/nnano.2006.54} {\bibfield  {journal}
  {\bibinfo  {journal} {Nature Nanotechnology}\ }\textbf {\bibinfo {volume}
  {1}},\ \bibinfo {pages} {53} (\bibinfo {year} {2006})}\BibitemShut {NoStop}%
\bibitem [{\citenamefont {Larsen}\ \emph {et~al.}(2015)\citenamefont {Larsen},
  \citenamefont {Petersson}, \citenamefont {Kuemmeth}, \citenamefont
  {Jespersen}, \citenamefont {Krogstrup}, \citenamefont {Nyg{\aa}rd},\ and\
  \citenamefont {Marcus}}]{larsen_semiconductor-nanowire-based_2015}%
  \BibitemOpen
  \bibfield  {author} {\bibinfo {author} {\bibfnamefont {T.~W.}\ \bibnamefont
  {Larsen}}, \bibinfo {author} {\bibfnamefont {K.~D.}\ \bibnamefont
  {Petersson}}, \bibinfo {author} {\bibfnamefont {F.}~\bibnamefont {Kuemmeth}},
  \bibinfo {author} {\bibfnamefont {T.~S.}\ \bibnamefont {Jespersen}}, \bibinfo
  {author} {\bibfnamefont {P.}~\bibnamefont {Krogstrup}}, \bibinfo {author}
  {\bibfnamefont {J.}~\bibnamefont {Nyg{\aa}rd}}, \ and\ \bibinfo {author}
  {\bibfnamefont {C.~M.}\ \bibnamefont {Marcus}},\ }\href {\doibase
  10.1103/PhysRevLett.115.127001} {\bibfield  {journal} {\bibinfo  {journal}
  {Physical Review Letters}\ }\textbf {\bibinfo {volume} {115}} (\bibinfo
  {year} {2015}),\ 10.1103/PhysRevLett.115.127001}\BibitemShut {NoStop}%
\bibitem [{\citenamefont {{de Lange}}\ \emph {et~al.}(2015)\citenamefont {{de
  Lange}}, \citenamefont {{van Heck}}, \citenamefont {Bruno}, \citenamefont
  {{van Woerkom}}, \citenamefont {Geresdi}, \citenamefont {Plissard},
  \citenamefont {Bakkers}, \citenamefont {Akhmerov},\ and\ \citenamefont
  {DiCarlo}}]{de_lange_realization_2015}%
  \BibitemOpen
  \bibfield  {author} {\bibinfo {author} {\bibfnamefont {G.}~\bibnamefont {{de
  Lange}}}, \bibinfo {author} {\bibfnamefont {B.}~\bibnamefont {{van Heck}}},
  \bibinfo {author} {\bibfnamefont {A.}~\bibnamefont {Bruno}}, \bibinfo
  {author} {\bibfnamefont {D.~J.}\ \bibnamefont {{van Woerkom}}}, \bibinfo
  {author} {\bibfnamefont {A.}~\bibnamefont {Geresdi}}, \bibinfo {author}
  {\bibfnamefont {S.~R.}\ \bibnamefont {Plissard}}, \bibinfo {author}
  {\bibfnamefont {E.~P. A.~M.}\ \bibnamefont {Bakkers}}, \bibinfo {author}
  {\bibfnamefont {A.~R.}\ \bibnamefont {Akhmerov}}, \ and\ \bibinfo {author}
  {\bibfnamefont {L.}~\bibnamefont {DiCarlo}},\ }\href {\doibase
  10.1103/PhysRevLett.115.127002} {\bibfield  {journal} {\bibinfo  {journal}
  {Physical Review Letters}\ }\textbf {\bibinfo {volume} {115}} (\bibinfo
  {year} {2015}),\ 10.1103/PhysRevLett.115.127002}\BibitemShut {NoStop}%
\end{thebibliography}%

\clearpage
\onecolumngrid

\section*{Supplementary material}

\subsection*{DC contact measurements}

\begin{figure}[th!]
\begin{center}
\includegraphics[width=\columnwidth]{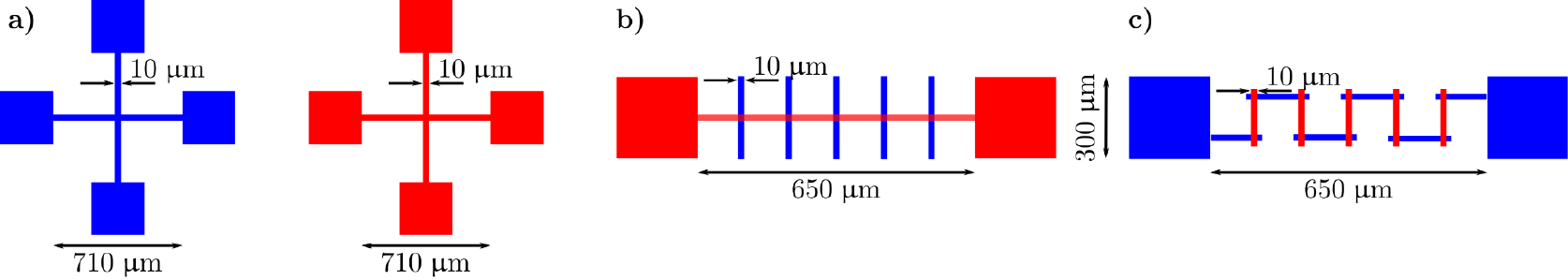}
\caption{Schematics of the four test patterns measured to extract the DC contact resistance $R_{\mathrm{c}}$. Blue color corresponds to the first lithographic layer, red to the second. \textbf{a)} A four probe measurement determines the sheet resistance $R_{\mathrm{s}}^{\left(1,2\right)}$ of the first and second aluminum thin film, respectively. \textbf{b)} We quantify the contribution of the edge resistance $R_{\mathrm{e}}$ to the total resistance by determining the total resistance of the sketched structure through a four probe measurement and application of Eq.\eqref{eq:Re}. \textbf{c)} For the extraction of the contact resistance $R_{\mathrm{c}}$ we measure the DC resistance of a design with ten ion milled contacts in series and disentangle $R_{\mathrm{c}}$ from the total measured resistance by using Eq.\eqref{eq:Rc}.}
\label{supfig:1}
\end{center}
\end{figure}

We identify two contributions to the resistance between two metal layers: an edge resistance $R_{\mathrm{e}}$ resulting from the reduced cross-section of the metal layer at the transition to the overlap contact, and a contact resistance $R_{\mathrm{c}}$ between two aluminum layers fabricated in separate lithographic steps. 
In order to be able to extract $R_{\mathrm{c}}$, we perform a four probe measurement on a set of four test patterns. 
Figure~\ref{supfig:1} shows schematic drawings of the four patterns, where layer 1 is drawn in blue and layer 2 is drawn in red. At first, we measure the sheet resistance of both aluminum thin films $R_{\mathrm{s}}^{\left(1,2\right)}$ (see Fig.~\ref{supfig:1}a). Here, the superscript corresponds to the respective number of the layer. Multiple measurements show very similar values for $R_{\mathrm{s}}^{\left(1,2\right)} = \unit{(1.2 \pm 0.1)}{\ohm\per} \Box$ of both films.
Measuring the DC resistance $R_{\mathrm{meas}}$ of the structure sketched in Fig.~\ref{supfig:1}b allows to calculate the edge resistance $R_{\mathrm{e}}$ by the following equation,

\begin{equation}
R_{\mathrm{e}} = \frac{1}{10} \left(R_{\mathrm{meas}} - R_{\mathrm{s}}^{\left(2\right)} N_{\mathrm{s}}^{\left(2\right)}\right), 
\label{eq:Re}
\end{equation}

where $N_{\mathrm{s}}^{(2)}$ is the number of squares of the red layer. From measurements on different samples we extract an upper bound for the edge resistance of $R_{\mathrm{e}} \leq \unit{0.25}{\ohm}$. 
Knowing $R_{\mathrm{s}}^{(1,2)}$ and $R_{\mathrm{e}}$ allows to calculate $R_{\mathrm{c}}$ with the equation below,

\begin{equation}
R_{\mathrm{c}} = \frac{1}{10}\left(R_{\mathrm{meas}} - R_{\mathrm{s}}^{\left(1\right)} N_{\mathrm{s}}^{\left(1\right)} - R_{\mathrm{s}}^{\left(2\right)} N_{\mathrm{s}}^{\left(2\right)} - 20 R_{\mathrm{e}}\right).
\label{eq:Rc}
\end{equation}

\subsection*{Calculation of the residual resistance}

Based on the average $\overline{Q_{\mathrm{i}}} = 1.5 \cdot 10^6$ of samples with argon ion milled contacts and a bridge (see Fig.~\ref{fig:3}b), we extract an upper bound on the residual resistance of the contacts. 
For the calculation we assume that all dissipation is caused by inductive loss, since the bridge and contacts are at a position of small electric and high magnetic field. The quality factor $Q$ of a lossy inductor is defined as $Q = \omega \cdot L / R$. Here, $\omega$ corresponds to the frequency, $L$ to the total inductance, and $R$ to the resistance of the inductor. 

We estimate the total inductance $L \approx \unit{33}{\nano\henry}$ by calculating the geometric inductance $L_{\mathrm{geo}}$ of a microstrip with a width of \unit{10}{\micro\meter} and a total length of \unit{1.6}{\centi\meter}, and taking into account the kinetic inductance fraction $\alpha = 0.14$ (see Fig.~\ref{fig:4}).
For a frequency $\omega = \unit{4.5}{\giga\hertz}$ we calculate the residual resistance to be
\begin{equation}
R = \dfrac{\omega L}{\overline{Q_{\mathrm{i}}}} = \unit{0.62}{\milli\ohm}.
\end{equation}

Since all resonators with a bridge have two contacts, we divide the total resistance by two and then multiply by the area of one argon ion milled contact. This gives a residual resistance of one overlap contact of,
\begin{equation}
R_{\mathrm{residual}} = \dfrac{\unit{0.62}{\milli\ohm}}{2} \unit{100}{\micro\meter\squared} = \unit{31}{\milli\ohm\cdot\micro\meter\squared},
\end{equation}

which we round up to $\unit{50}{\milli\ohm\cdot\micro\meter\squared}$ to quote a conservative upper bound in the main text.

\subsection*{Parameter list of resonators}

\begin{table}[!h]
\begin{center}
\caption{Extensive parameter list of the four resonators on each sample. The powers indicated in the column of $Q_{\mathrm{i}}$ correspond to the readout power reaching the sample holder input port, taking into account the total estimated attenuation of our measurement setup.}
\begin{tabular}{c|c|c||c|c|c|c|c|c}
\hline \hline
\multicolumn{3}{c||}{} & \multicolumn{6}{c}{$Q_{\mathrm{i}}~(10^6)$} \\ 
sample & $f_{\mathrm{r}}~(\unit{}{\giga\hertz})$ & $Q_{\mathrm{c}}~(10^6)$ & -120 dBm & -130 dBm & -140 dBm & -150 dBm & -160 dBm & -170 dBm \\ \hline
\multirow{5}{*}{A} 	& 4.4774 & 9.7 & 8.3 & 6.0 & 4.2 & 3.4 & 2.7 & 2.3 \\ 
					& 4.5503 & 4.8 & 7.4 & 5.8 & 4.3 & 3.0 & 1.3 & 0.5 \\ 
					& 4.6257 & 0.6 & \multicolumn{6}{l}{The amplitude signal shows an asymmetric response, which is not} \\ 
					&  &     	   & \multicolumn{6}{l}{captured in our model. Therefore, no values for $Q_{\mathrm{i}}$ can be extracted} \\ 
 					& 4.7035 & 0.9 & 3.2 & 2.7 & 2.3 & 1.9 & 1.6 & 0.8 \\ \hline
 
\multirow{4}{*}{B} 	& 4.5128 & 11.9  & 13.7 & 10.4 & 6.4 & 2.8 & 1.3 & 0.9 \\ 
					& 4.5872 & 2.6 & 8.7 & 7.7 & 5.5 & 3.6 & 2.0 & 1.5 \\ 
					& 4.6649 & 0.7 & \multicolumn{2}{c}{Not converging\footnote{The amplitude response shows a slight asymmetry, that dominates at higher readout powers. For smaller powers this only influences the shape of the amplitude response of the resonator marginally and is therefore neglected.}} & 4.1 & 2.5 &  1.6 & 1.2 \\ 
 					& 4.7474 & 1.2 & 4.9 & 4.0 & 3.2 & 2.3 & 1.9 & 1.0 \\ \hline
 					
\multirow{4}{*}{C} 	& 4.4811 & 7.3 & 11.3 & 8.8 & 6.6 & 4.2 & 3.4 & 1.6 \\ 
					& 4.5586 & 1.4 & 11.7 & 9.5 & 7.0 & 4.0 & 3.4 & 1.4 \\ 
					& 4.6370 & 0.5 & 21.9 & 14.6 & 9.0 & 5.4 & 2.9 & 2.4 \\ 
 					& \multicolumn{8}{c}{The meander of the fourth resonator was interrupted during fabrication.} \\ \hline
 					
\multirow{5}{*}{D} 	& 4.5310 & 30.5 & 8.6 & 6.4 & 3.0 & 1.1 & \multicolumn{2}{c}{Not converging\footnote{Due to the very weak coupling the resonator response exhibits a very poor signal to noise ratio at low readout powers}} \\ 
					& 4.6050 & 4.4 & 3.9 & 3.3 & 2.4 & 1.2 & 0.8 & 0.5 \\ 
					& 4.6839 & 0.2 & \multicolumn{6}{l}{The amplitude response shows a peak instead of a dip, which is not} \\ 
					&  &     	   & \multicolumn{6}{l}{captured in our model. Therefore, no values for $Q_{\mathrm{i}}$ can be extracted} \\ 
 					& 4.7646 & 0.9 & 1.5 & 1.3 & 1.2 & 1.0 & 0.8 & 0.5 \\ \hline\hline
 					
\end{tabular} 
\end{center}
\end{table}

\clearpage

\subsection*{Sample shielding}

\begin{figure}[h!]
\begin{center}
\includegraphics[width=\columnwidth]{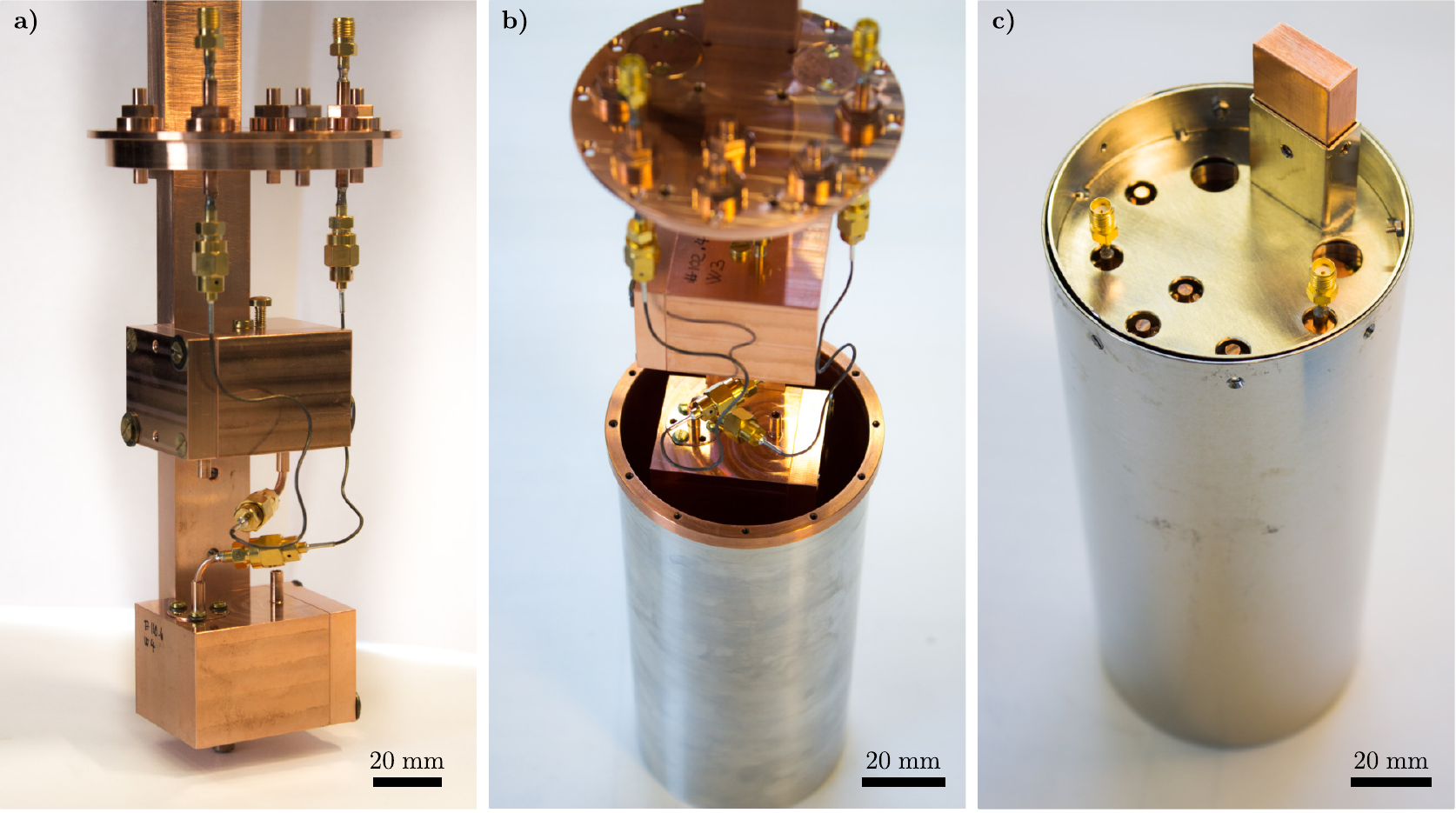}
\caption{Photographs of the sample shielding assembly. \textbf{a)} A central copper rod holds two 3D copper waveguide sample holders. Superconducting coaxial cables connect the ports of the waveguides to the feed through in the cap of the shielding barrel, minimizing signal loss for the reflection measurement. \textbf{b)} The barrel machined from a copper/aluminum sandwich encloses both waveguides and provides IR radiation and magnetic shielding. For additional IR absorption we cover the inside of the barrel with $\mathrm{Stycast}^{\mathrm{TM}}$ (not visible here). An aluminum cover is added to the copper cap of the barrel in a second step. \textbf{c)} The outermost layer of our sample holder assembly is a $\mu$-metal enclosure that provides additional magnetic shielding. We mount the entire assembly into a commercial dilution cryostat by inserting the copper rod into a specially designed copper holder, that is thermally anchored to the mixing chamber plate and clamps the rod, which provides thermal contact.}
\label{supfig:2}
\end{center}
\end{figure}

\end{document}